\newif\ifws
\ifws

\documentclass{ws-book9x6}
\usepackage{ws-book-thm}   
\usepackage{ws-book-har}   
\usepackage[pdfpagelabels=false]{hyperref}  
\makeindex

\usepackage{eepic}
\usepackage{xcolor}

\begin{document}

\titlepages                        

\chapter{Generalized event structures and probabilities}
\begin{center}
{Karl Svozil}\\
{\it Institute for Theoretical Physics, Vienna  University of Technology}\\
{\it  Wiedner Hauptstra\ss e 8-10/136, A-1040  Vienna, Austria}\\
{\it email: {svozil@tuwien.ac.at} homepage: {http://tph.tuwien.ac.at/\char`\~svozil} }
\end{center}


\else
\documentclass[%
 12pt,
 reprint,
  twocolumn,
 showpacs,
 showkeys,
 preprintnumbers,
 amsmath,amssymb,
 aps,
  pra,
  longbibliography,
 ]{revtex4-1}

\usepackage[breaklinks=true,colorlinks=true,anchorcolor=blue,citecolor=blue,filecolor=blue,menucolor=blue,pagecolor=blue,urlcolor=blue,linkcolor=blue]{hyperref}
\usepackage{url}

\usepackage{eepic}
\RequirePackage{times}
\RequirePackage{mathptm}
\usepackage{xcolor}

\begin{document}

\title{Generalized event structures and probabilities}


\author{Karl Svozil}
\affiliation{Institute for Theoretical Physics, Vienna
    University of Technology, Wiedner Hauptstra\ss e 8-10/136, A-1040
    Vienna, Austria}
\email{svozil@tuwien.ac.at} \homepage[]{http://tph.tuwien.ac.at/~svozil}


\begin{abstract}
For the classical mind, quantum mechanics is boggling enough; nevertheless more bizarre behavior could be imagined, thereby concentrating on propositional structures (empirical logics) that transcend the quantum domain. One can also consistently suppose predictions and probabilities which are neither classical nor quantum, but which are subject to subclassicality; that is, the additivity of probabilities for mutually exclusive, co-measurable observables, as formalized by admissibility rules and frame functions.
\end{abstract}

\maketitle
\fi

\section{Specker's oracle}

In his first, programmatic, article on quantum logic Ernst Specker
--
one of his sermons is preserved in his
{\it Selecta}~\cite[pp.~321-323]{specker-ges}
--
considered a parable~\cite{specker-60} which can be easily translated into the following oracle:
imagine that there are three boxes on a table, each of which either contains a gem or does not. Your task is to correctly choose two of the boxes that will either both be empty or both contain a gem when opened.

Note that, according to combinatorics (or, more generally, Ramsey theory), for all classical states
there always exist two such boxes among the three boxes satisfying the above property
of being ``both empty or both filled.''

After you place your guess the two boxes whose content you have attempted to predict are opened;
the third box remains closed.
In Specker's malign oracle scenario it turns out that you always fail:
no matter how often you try and what you choose to forecast,
the boxes you have predicted as both being empty or both being full
always have mixed content -- one box is always filled and the other one always empty.
That is, phenomenologically, or, if you like, epistemically, Specker's oracle
is defined by the following behavior:
if $e$ and $f$ denote the empty and the filled state, respectively, and $\ast$ stands for
the third (unopened) box,
then one of the following six configurations are rendered:
$ef\ast$,
$fe\ast$,
$e\ast f$,
$f\ast e$,
$\ast ef$, or
$\ast fe$.

Is such a Specker oracle realizable in Nature?
Intuition tends to negate this.
Because, more formally, per box
there are two classical states $e$ and $f$, and thus $2^3$ such classical ``ontological''
configurations or classical three-box
states,
namely
$eee$,
$eef$,
$efe$,
$fee$,
$eff$,
$fef$,
$ffe$,
and
$fff$, which can be grouped into four classes: those extreme cases with all the boxes filled and empty, those
with two empty and one filled boxes, and those with two filled and one empty boxes.
These can be represented by the four-partitioning (into equivalence classes with respect to the number of filled and empty boxes)
of the set of all states
$
\{
\{eee
\},
\{fff
\},
\{eef,efe.fee
\},
\{eff,fef,eff
\}
\}
$.

Now, on closer inspection, in any unbiased prediction (or unbiased preparation)
scenario there is an ever decreasing chance that you will not hit the right prediction eventually,
because for all eight possible configurations there always is at least one right prediction
(either two empty or two full boxes).

Of course, if I am in command of the preparation process,
and if you and me chose to conspire in such a way that I always choose to prepare, say, either
$eee$ or
$eef$ or
$efe$ or
$fee$,
and you always choose to predict $ff$, than you will never win.
But such a scenario is hilariously biased.
Also with adaptive, that is, {\it a posteriori,} preparation {\em after}
the prediction, the Specker parable is realizable --
in hindsight I can always ruin your prediction.
But if you allow no restrictions on predictions (or preparations),
and no {\it a posteriori} manipulation,
there are no classical means to realize Specker's oracle.

Can this system be realized quantum mechanically?
That is, can one find a quantum state and projection measurements rendering that kind of performance?
I guess (but have no proof of it) not, because
in any finite dimensional Hilbert space
the associated empirical logic~\cite{v-neumann-49,birkhoff-36} is a merging through identifying common elements, called a {\em pasting}~\cite{nav:91},
of (possibly a continuum of) Boolean subalgebras with a finite number of atoms
or, used synonymously, contexts~\cite{svozil-2008-ql,2014-nobit}.
And any subalgebra, according to the premises of
Gleason's theorem~\cite{Gleason,r:dvur-93,pitowsky:218,Peres-expTest-Glea},
in terms of probability theory, is classically Boolean.

As has already been pointed out by Specker, the phenomenology of the oracle
suggests, that $e_i \rightarrow f_j$, and, conversely $f_i \rightarrow e_j$ for different Boxes $i,j \in \{1,2,3\}$,
that is, ``the first opened box always contains the complement of the second opened box'';
and otherwise -- that is, by disregarding the third (unopened) box -- they are classical.
Thus one could say that the contents of the two opened boxes represent the two atoms of a Boolean subalgebra $2^2$.
There are three such subalgebras associated with opening two of three boxes, namely
$(1,2)$,
$(1,3)$, and
$(2,3)$
which need to be pasted into the propositional structure at hand; in the quantum case this is quantum logic.

This can be imagined in two ways, by interpreting the situation as follows:
(i) The first option would be to attempt to paste or ``isomorphically bundle''
the three subalgebras $2^2$ into a three-atomic subalgebra $2^3$.
Clearly this attempt is futile, since this would imply transitivity, and thus yield a complete contradiction, by,
say $e_1 \rightarrow f_2 \rightarrow e_3 \rightarrow f_1$.
(ii) The second option would circumvent transitivity by means of complementarity (as argued originally by Specker),
through a {\em horizontal pasting} of the three Boolean algebras, amounting to a logic of the Chinese lantern form ${\rm MO}_3$.
This is a common quantum logic rendered, for instance, by spin-$\frac{1}{2}$ measurements along different spatial directions;
as well as by the quasi-classical partition logics~\cite{svozil-2001-eua}
of automata and generalized urn models~\cite{wright}.
But clearly, such a logic does not deal with the three boxes of Specker's oracle equally;
rather the third, unopened box
could be considered as a ``space holder'' or ``indicator'' labeling the associated context.

Within such a context one could, for example, attempt to consider a general wave function in eight dimensional Hilbert space
$\vert \Psi \rangle =\sum_{i,j,k \in \{e,f\}}\alpha_{ijk} \vert ijk \rangle$,
geometrically representable by $\vert e\rangle \equiv (1,0)$  and $\vert f \rangle \equiv (0,1)$,
and thus $\vert \Psi \rangle \equiv \left(\alpha_{eee},\alpha_{eef},\ldots ,\alpha_{fff} \right)$.
All three measurements (i.e. projections onto $\vert ijk \rangle$) commute;
so one can open the boxes ``independently.''
By listing all the associated ``unbiased'' measurement scenarios (including partial traces over the third box),
there is no quantum way one could end up with the type of behavior one expects from Specker's oracle.
Ultimately, because a general quantum state is a coherent superposition of classical states,
one cannot ``break outside'' this extended classical domain.

So, I guess,
if one insists on treating all the three boxes involved in Specker's oracle equally,
this device requires supernatural means.
And yet it is imaginable; and that is the beauty of it.

\section{Observables unrealizable by quantum means}

In what follows we shall enumerate, as a kind of continuation of Specker's oracle,
hypothetical ``weird'' propositional structures,
in particular,
certain anecdotal ``zoo of collections of observables''
constructed by pastings of contexts
(or, used synonymously, blocks, subalgebras) containing ``very few'' atoms.
We shall compare them to logical structures associated with
very low-dimensional quantum Hilbert spaces.
(Actually,
the dimensions dealt with will never exceed the number of fingers on one hand.)

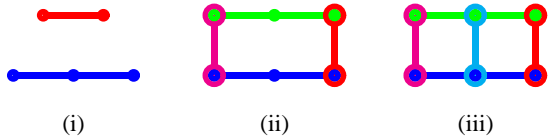
\begin{figure}
\begin{center}
\begin{tabular}{ccccc}
\unitlength 0.8mm 
\allinethickness{2.1pt}
\ifx\plotpoint\undefined\newsavebox{\plotpoint}\fi 
\begin{picture}(21,21)(0,0)
\put(0,0){\color{blue}\circle{1.2}}
\put(10,0){\color{blue}\circle{1.2}}
\put(20,0){\color{blue}\circle{1.2}}
\put(0,0){\color{blue}\line(1,0){20}}
\put(5,10){\color{red}\circle{1.2}}
\put(15,10){\color{red}\circle{1.2}}
\put(5,10){\color{red}\line(1,0){10}}
\end{picture}
&&
\unitlength 0.8mm 
\allinethickness{2.1pt}
\ifx\plotpoint\undefined\newsavebox{\plotpoint}\fi 
\begin{picture}(21,21)(0,0)
\put(0,0){\color{blue}\line(1,0){20}}
\put(20,0){\color{red}\line(0,1){10}}
\put(0,10){\color{green}\line(1,0){20}}
\put(0,10){\color{magenta}\line(0,-1){10}}
\put(0,0){\color{blue}\circle{1.2}}
\put(0,0){\color{magenta}\circle{3}}
\put(0,10){\color{green}\circle{1.2}}
\put(0,10){\color{magenta}\circle{3}}
\put(10,0){\color{blue}\circle{1.2}}
\put(10,10){\color{green}\circle{1.2}}
\put(20,0){\color{blue}\circle{1.2}}
\put(20,0){\color{red}\circle{3}}
\put(20,10){\color{green}\circle{1.2}}
\put(20,10){\color{red}\circle{3}}
\end{picture}
&&
\unitlength 0.8mm 
\allinethickness{2.1pt}
\ifx\plotpoint\undefined\newsavebox{\plotpoint}\fi 
\begin{picture}(21,21)(0,0)
\put(0,0){\color{blue}\line(1,0){20}}
\put(20,0){\color{red}\line(0,1){10}}
\put(0,10){\color{green}\line(1,0){20}}
\put(0,10){\color{magenta}\line(0,-1){10}}
\put(10,0){\color{cyan}\line(0,1){10}}
\put(0,0){\color{blue}\circle{1.2}}
\put(0,0){\color{magenta}\circle{3}}
\put(0,10){\color{green}\circle{1.2}}
\put(0,10){\color{magenta}\circle{3}}
\put(10,0){\color{blue}\circle{1.2}}
\put(10,0){\color{cyan}\circle{3}}
\put(10,10){\color{green}\circle{1.2}}
\put(10,10){\color{cyan}\circle{3}}
\put(20,0){\color{blue}\circle{1.2}}
\put(20,0){\color{red}\circle{3}}
\put(20,10){\color{green}\circle{1.2}}
\put(20,10){\color{red}\circle{3}}
\end{picture}
\\
$\;$\\
(i)&$\qquad$&(ii)&$\qquad$&(iii)
\end{tabular}
\end{center}
\caption{(Color online) Orthogonality diagrams with mixed two- and three-atomic contexts, drawn in different colors.
\label{2015-s-f1}}
\end{figure}

It is not too difficult to sketch propositional structures which are not realizable by any known physical device.
Take, for instance, the collection of observables whose  Greechie or, by another wording, orthogonality diagram~\cite{greechie:71}
is sketched in Fig.~\ref{2015-s-f1}.
In Hilbert space realizations,
the straight lines or smooth curves depicting contexts represent orthogonal bases,
and points on these straight lines or smooth curves represent elements of these bases;
that is, two points being orthogonal
if and only if they are on the same these straight line or smooth curve.
From dimension three onwards, bases can intertwine~\cite{Gleason} by possessing common elements.

The propositional structure depicted in Fig.~\ref{2015-s-f1} consists of four contexts of mixed type;
that is, the contexts involved have two and three atoms.
No such mixed type phenomenology occurs in Nature;
on the contrary, regardless of the quantized system
the number of (mutually exclusive) physical outcomes, reflected by the dimension of the
associated Hilbert space, always remains the same.

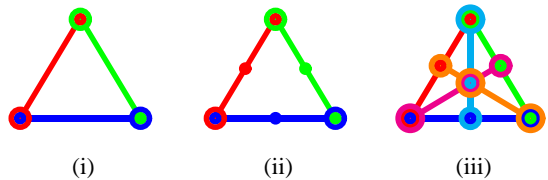
\begin{figure}
\begin{center}
\begin{tabular}{ccccc}
\unitlength 0.8mm 
\allinethickness{2.1pt}
\ifx\plotpoint\undefined\newsavebox{\plotpoint}\fi 
\begin{picture}(21,25)(0,0)
\put(0,0){\color{blue}\line(1,0){20}}
\put(0,0){\color{red}\line(3,5){10}}
\put(20,0){\color{green}\line(-3,5){10}}
\put(0,0){\color{blue}\circle{1.2}}
\put(0,0){\color{red}\circle{3}}
\put(20,0){\color{green}\circle{1.2}}
\put(20,0){\color{blue}\circle{3}}
\put(10,16.5){\color{red}\circle{1.2}}
\put(10,16.5){\color{green}\circle{3}}
\end{picture}
&&
\unitlength 0.8mm 
\allinethickness{2.1pt}
\ifx\plotpoint\undefined\newsavebox{\plotpoint}\fi 
\begin{picture}(21,25)(0,0)
\put(0,0){\color{blue}\line(1,0){20}}
\put(0,0){\color{red}\line(3,5){10}}
\put(20,0){\color{green}\line(-3,5){10}}
\put(0,0){\color{blue}\circle{1.2}}
\put(0,0){\color{red}\circle{3}}
\put(20,0){\color{green}\circle{1.2}}
\put(20,0){\color{blue}\circle{3}}
\put(10,16.5){\color{red}\circle{1.2}}
\put(10,16.5){\color{green}\circle{3}}
\put(5,8.25){\color{red}\circle{1.2}}
\put(15,8.25){\color{green}\circle{1.2}}
\put(10,0){\color{blue}\circle{1.2}}
\end{picture}
&&
\unitlength 0.8mm 
\allinethickness{2.2pt}
\ifx\plotpoint\undefined\newsavebox{\plotpoint}\fi 
\begin{picture}(21,25)(0,0)
\put(0,0){\color{blue}\line(1,0){20}}
\put(0,0){\color{red}\line(3,5){10}}
\put(20,0){\color{green}\line(-3,5){10}}
\put(10,0){\color{cyan}\line(0,1){16.5}}
\put(5.3,8.75){\color{orange}\line(5,-3){15}}
\put(14.7,8.75){\color{magenta}\line(-5,-3){15}}
\put(0,0){\color{blue}\circle{1.2}}
\put(0,0){\color{red}\circle{3}}
\put(0,0){\color{magenta}\circle{4}}
\put(20,0){\color{green}\circle{1.2}}
\put(20,0){\color{blue}\circle{3}}
\put(20,0){\color{orange}\circle{4}}
\put(10,16.5){\color{red}\circle{1.2}}
\put(10,16.5){\color{green}\circle{3}}
\put(10,16.5){\color{cyan}\circle{4}}
\put(5,8.75){\color{red}\circle{1.2}}
\put(5,8.75){\color{orange}\circle{3}}
\put(15,8.75){\color{green}\circle{1.2}}
\put(15,8.75){\color{magenta}\circle{3}}
\put(10,0){\color{blue}\circle{1.2}}
\put(10,0){\color{cyan}\circle{3}}
\put(10,5.9){\color{cyan}\circle{1.2}}
\put(10,5.9){\color{magenta}\circle{3}}
\put(10,5.9){\color{orange}\circle{4}}
\end{picture}
\\
$\;$\\
(i)&$\qquad$&(ii)&$\qquad$&(iii)
\end{tabular}
\end{center}
\caption{(Color online) Orthogonality diagrams representing tight triangular pastings of two- and three-atomic contexts.
\label{2015-s-f2}}
\end{figure}

You may now say that this was an easy and almost trivial cheat;
but what about the triangular shaped propositional structures depicted in
Fig.~\ref{2015-s-f2}?
They surely look inconspicuous,
yet none of them has a representation as a quantum logic;
simply because they have no realization in two- and
three-dimensional Hilbert space:
The propositional structure depicted in Fig.~\ref{2015-s-f2}(i)
has too tightly intertwining contexts, which would mean that
two different orthogonal bases in two-dimensional Hilbert space
can have an element in common
(which they cannot have, except when the bases are identical).
By a similar argument,
the propositional structure depicted in Fig.~\ref{2015-s-f2}(ii) has
``too tightly intertwined'' contexs to be representable in
three-dimensional Hilbert space:
in dimension three, for two non-identical but intertwined orthogonal bases with one common vector
(if they have two common elements they would have to be identical)
it is impossible to ``shuffle'' the remaining vectors around such that
at least one remaining vector from one basis is orthogonal to at least one remaining vector from the other basis.
From an algebraic point of view all these propositional structures are not realizable quantum mechanically,
because they contain loops of order three~\cite{kalmbach-83,beran,pulmannova-91}.

Indeed, for reasons that will be explicated later,
the propositional structure depicted in Fig.~\ref{2015-s-f2}(i) has no two-valued
(admissible~\cite{2012-incomput-proofsCJ,PhysRevA.89.032109,2015-AnalyticKS}) state
equivalent to a frame function~\cite{Gleason};
a fact that can be seen by
ascribing one element a ``1,'' forcing the remaining two to be ``0.'' (There cannot be only zeroes in a context.)
This means that it is no quasi classical partition logic.
The logic depicted in Fig.~\ref{2015-s-f2}(ii) has sufficiently many (indeed four)
two-valued measures
to be representable by a partition logic~\cite{2010-qchocolate}.
The propositional structure depicted in Fig.~\ref{2015-s-f2}(iii)
is too tightly interlinked to be representable by a partition logic -- it allows only one two-valued state.

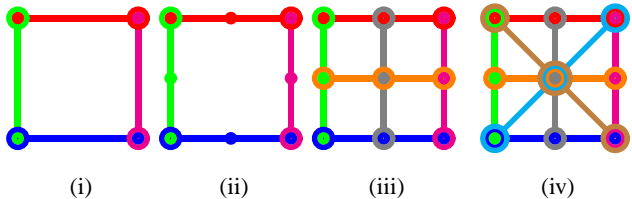
\begin{figure}
\begin{center}
\begin{tabular}{ccccccc}
\unitlength 0.8mm 
\allinethickness{2.1pt}
\ifx\plotpoint\undefined\newsavebox{\plotpoint}\fi 
\begin{picture}(21,21)(0,0)
\put(0,0){\color{green}\line(0,1){20}}
\put(0,0){\color{blue}\line(1,0){20}}
\put(20,0){\color{magenta}\line(0,1){20}}
\put(0,20){\color{red}\line(1,0){20}}
\put(0,0){\color{green}\circle{1.2}}
\put(0,0){\color{blue}\circle{3}}
\put(20,0){\color{blue}\circle{1.2}}
\put(20,0){\color{magenta}\circle{3}}
\put(0,20){\color{red}\circle{1.2}}
\put(0,20){\color{green}\circle{3}}
\put(20,20){\color{magenta}\circle{1.2}}
\put(20,20){\color{red}\circle{3}}
\end{picture}
&&
\unitlength 0.8mm 
\allinethickness{2.1pt}
\ifx\plotpoint\undefined\newsavebox{\plotpoint}\fi 
\begin{picture}(21,21)(0,0)
\put(0,0){\color{green}\line(0,1){20}}
\put(0,0){\color{blue}\line(1,0){20}}
\put(20,0){\color{magenta}\line(0,1){20}}
\put(0,20){\color{red}\line(1,0){20}}
\put(0,0){\color{green}\circle{1.2}}
\put(0,0){\color{blue}\circle{3}}
\put(10,0){\color{blue}\circle{1.2}}
\put(20,0){\color{blue}\circle{1.2}}
\put(20,0){\color{magenta}\circle{3}}
\put(20,10){\color{magenta}\circle{1.2}}
\put(0,20){\color{red}\circle{1.2}}
\put(0,20){\color{green}\circle{3}}
\put(0,10){\color{green}\circle{1.2}}
\put(20,20){\color{magenta}\circle{1.2}}
\put(20,20){\color{red}\circle{3}}
\put(10,20){\color{red}\circle{1.2}}
\end{picture}
&&
\unitlength 0.8mm 
\allinethickness{2.1pt}
\ifx\plotpoint\undefined\newsavebox{\plotpoint}\fi 
\begin{picture}(21,21)(0,0)
\put(0,0){\color{green}\line(0,1){20}}
\put(0,0){\color{blue}\line(1,0){20}}
\put(20,0){\color{magenta}\line(0,1){20}}
\put(0,20){\color{red}\line(1,0){20}}
\put(0,10){\color{orange}\line(1,0){20}}
\put(10,0){\color{gray}\line(0,1){20}}
\put(0,0){\color{green}\circle{1.2}}
\put(0,0){\color{blue}\circle{3}}
\put(10,0){\color{blue}\circle{1.2}}
\put(10,0){\color{gray}\circle{3}}
\put(20,0){\color{blue}\circle{1.2}}
\put(20,0){\color{magenta}\circle{3}}
\put(20,10){\color{magenta}\circle{1.2}}
\put(20,10){\color{orange}\circle{3}}
\put(0,20){\color{red}\circle{1.2}}
\put(0,20){\color{green}\circle{3}}
\put(0,10){\color{green}\circle{1.2}}
\put(0,10){\color{orange}\circle{3}}
\put(20,20){\color{magenta}\circle{1.2}}
\put(20,20){\color{red}\circle{3}}
\put(10,20){\color{red}\circle{1.2}}
\put(10,20){\color{gray}\circle{3}}
\put(10,10){\color{gray}\circle{1.2}}
\put(10,10){\color{orange}\circle{3}}
\end{picture}
&&
\unitlength 0.8mm 
\allinethickness{2.1pt}
\ifx\plotpoint\undefined\newsavebox{\plotpoint}\fi 
\begin{picture}(21,21)(0,0)
\put(0,0){\color{green}\line(0,1){20}}
\put(0,0){\color{blue}\line(1,0){20}}
\put(20,0){\color{magenta}\line(0,1){20}}
\put(0,20){\color{red}\line(1,0){20}}
\put(0,10){\color{orange}\line(1,0){20}}
\put(0,0){\color{cyan}\line(1,1){20}}
\put(20,0){\color{brown}\line(-1,1){20}}
\put(10,0){\color{gray}\line(0,1){20}}
\put(0,0){\color{green}\circle{1.2}}
\put(0,0){\color{blue}\circle{3}}
\put(0,0){\color{cyan}\circle{4}}
\put(10,0){\color{blue}\circle{1.2}}
\put(10,0){\color{gray}\circle{3}}
\put(20,0){\color{blue}\circle{1.2}}
\put(20,0){\color{magenta}\circle{3}}
\put(20,0){\color{brown}\circle{4}}
\put(20,10){\color{magenta}\circle{1.2}}
\put(20,10){\color{orange}\circle{3}}
\put(0,20){\color{red}\circle{1.2}}
\put(0,20){\color{green}\circle{3}}
\put(0,20){\color{brown}\circle{4}}
\put(0,10){\color{green}\circle{1.2}}
\put(0,10){\color{orange}\circle{3}}
\put(20,20){\color{magenta}\circle{1.2}}
\put(20,20){\color{red}\circle{3}}
\put(20,20){\color{cyan}\circle{4}}
\put(10,20){\color{red}\circle{1.2}}
\put(10,20){\color{gray}\circle{3}}
\put(10,10){\color{gray}\circle{1.2}}
\put(10,10){\color{orange}\circle{3}}
\put(10,10){\color{cyan}\circle{4}}
\put(10,10){\color{brown}\circle{5}}
\end{picture}
\\
$\qquad$\\
(i)&$\;$&(ii)&$\;$&(iii)&$\quad$&(iv)
\end{tabular}
\end{center}
\caption{(Color online) Orthogonality diagrams representing tight square type pastings of two- and three-atomic contexts.
\label{2015-s-f3}}
\end{figure}

In a similar manner one could go on and consider orthogonality diagrams of the ``square'' type, such as the ones depicted in
Fig.~\ref{2015-s-f3}.
All these propositional structures are not realizable quantum mechanically, because they contain loops of order four~\cite{kalmbach-83,beran,pulmannova-91}.
The propositional structure in Fig.~\ref{2015-s-f3}(i) has two two-valued measures,
but the union of them is not ``full''
because it cannot separate opposite atoms.
Figs.~\ref{2015-s-f3}(ii) as well as (iii) represent propositional structures with
``sufficiently many'' two-valued measures (e.g. separating two arbitrary atoms by different values),
which are representable as partition
(and, in particular, as generalized urn and automaton) logics.
Actually, the number of two-valued measures for the propositional structures in Figs.~\ref{2015-s-f3}(i) as well as (iii)
can be found by counting the number of permutations, or permutation matrices: these are $2!$ and $3!$, respectively.
Because of the too tightly intertwined contexts the propositional structure in Fig.~\ref{2015-s-f3}(iv)
has no two-valued state.

Let us now come back to the collection of observables represented in Fig.~\ref{2015-s-f1}.
Are they in some form realizable, maybe even in ways ``beyond'' quantum realizability?
Again, as long as there are ``sufficiently many'' two-valued measures~\cite{wright:pent},
partition logics
as well as their generalized urn and automaton models~\cite{svozil-2001-eua}
are capable of reproducing these phenomenological schemes.
One construction yielding the pasting described in Fig.~\ref{2015-s-f1}(ii)
would involve a four color (associated with the four contexts) scheme;
with three symbols
``$+$,''
``$-$,'' and
``$0$''
in two colors
representing the Boolean algebra $2^3$ of two contexts,
and with two symbols
``$+$,''  and
``$-$''
in two colors
representing the Boolean algebra $2^2$ of two contexts.
(I leave it to the Reader to find a concrete realization;
one systematic way would be the enumeration of all two-valued measures.)
Fig.~\ref{2015-s-f1}(iii) does not possess a quasi-classical {\it simulacrum}
in terms of a partition logic.
For the sake of a proof by contradiction~\cite{greechie:71}, suppose there exist a two-valued state.
Any such two-valued state
needs to have exactly two 1s on the horizontal contexts, whereas it
needs to have exactly three 1s on the vertical contexts;
but both contexts yield (two- and three-atomic) partitions of the entire set of atoms;
thus implying $2=3$, which is clearly wrong.

So, in a sense, one could say that the collection of observables represented in Fig.~\ref{2015-s-f1}(iii)
is ``weirder'' than the ones represented in Figs.~\ref{2015-s-f1}(i)-(iii).

\section{Generalized probabilities beyond the quantum predictions}

When it comes to observables and probabilities there are two fundamental questions:
(i) Given a particular collection of observables; what sort of probability measures can this propositional structure support or entail~\cite{Pitowsky2003395,pitowsky-06}?
(ii) Conversely, given a particular probability measure;
which observables and what propositional structure can be associated with this probability~\cite{Hardy:2001jk,Hardy2003381}?
We shall mainly concentrate on the first question.

A {\it caveat} is in order: it might as well be that, from a certain perspective,
we might not be forced to ``leave'' or modify classical probability theory: for example,
quantum probabilities could be interpreted as classical {\em conditional} probabilities~\cite{Khrennikov-15},
where conditioning
is with respect to fixed experimental settings, in particular, with respect to the context measured.

\subsection{Subclassicality and frame functions}

In order to construct probability measures on non-Boolean propositional structures
which can be obtained by pasting together contexts
we shall adhere to the following assumption which we would like to call {\em subclassicality}:
{\em every context (i.e., Boolean subalgebra, block) is endowed with classical probabilities.}
In particular, any probability measure on its atoms is additive.
This is quite reasonable, because it is prudent to maintain the validity of classical probability theory on those classical
substructures of the propositional calculus that containing observables which are mutually co-measurable.

Subclassicality can be formalized by
frame functions in the context of Hilbert spaces~\cite{Gleason,r:dvur-93,pitowsky:218,Peres-expTest-Glea} as follows:
A frame function of unit weight for a separable Hilbert space $H$ is a
real-valued function $f$ defined on the (surface of the) unit sphere of $H$ such that if
$\{ e_i \}$ is an orthonormal basis of $H$ then $\sum_i f(e_i) = 1$.
This can be translated for pastings of contexts
by identifying the set of atoms $\{a_{i}\}$ in a particular context $C$
with the set of vectors in one basis, and by requireing that  $\sum_i f(a_i) = 1$ for all contexts $C$ involved.

For pastings of contexts on value definite systems of observables, admissibility,
which originally has been conceived as a formalization of ``partial value definiteness'' and
value indefiniteness~\cite{2012-incomput-proofsCJ,PhysRevA.89.032109,2015-AnalyticKS}
is essentially equivalent to the requirements imposed upon frame functions; that is, subclassicality.
Nevertheless, one could also request generalized admissibility rules as follows.
        Let $O$ be a set of atoms in a propositional structure,
and let $f: O \to [0,1]$ be a probability measure.
        Then $f$ is {\em admissible} if the following condition holds for every context $C$
of $ O $:
        for any $a\in C$ with $0\le f(a)\le 1$,  $\sum_i f(b_i)=1-f(a)$ for all $b_i\in C\setminus\{a\}$.
Likewise, for two-valued measures $v$ on value definite systems of observables,
admissibility~\cite{2012-incomput-proofsCJ,PhysRevA.89.032109,2015-AnalyticKS} can be
defined in analogy to frame functions:  for any context $C=\{a_1,\ldots,a_n\}$
of $ O $,
the two-valued measure on the atoms $a_1,\ldots,a_n$ has to add up to one; that is,
 $\sum_i v(a_i) = 1$.

For the sake of a (quasi-)  classical formalization,
define a {\em two-valued measure} (or, used synonymously, {\em valuation,} or {\em truth assignment})
$v$ on a single context $C=\{a_1,\ldots,a_n\}$
to acquire the value $v (a_i)=1$ on exactly one $a_i$, $1\le i\le n$ of the atoms of the context,
and the value zero on the remaining atoms $v (a_{j\neq i})=0$, $1\le j\le n$.
Any (quasi-)  classical probability measure, or, used synonymously, {\em state,} or
{\em non-negative frame function} $f$ (of weight one),
on this context can then be obtained by a convex combination of all $m$ two-valued measures;
that is,
\begin{equation}
\begin{split}
f =\sum_{1\le k\le m} \lambda_k v_k
\text{, with }  \\
1 =\sum_{1\le k\le m} \lambda_k
\text{, and }
\lambda_k  \ge 0.
\label{2015-s-e1}
\end{split}
\end{equation}
As far as classical physics is concerned, that is all there is -- the classical probabilities are just the
convex combinations of the $m$ two-valued measures on the Boolean algebras $2^m$.

This convex combination can be given a geometrical interpretation:
First encode every two-valued measure on $C$ as some $m$-tuple,
whereby the $i$'th component of the  $m$-tuple is identified
with the value $v(a_i)$ of that valuation on the $i$'th atom of the context $C$;
and then interpret the resulting set of $m$-tuples as the set of the vertices of a convex polytope.

By the Minkoswki-Weyl representation theorem~\cite[p.29]{ziegler},
every convex polytope has a dual (equivalent) description:
either as the convex hull of its extreme points (vertices);
or as the intersection of a finite number of half-spaces.
More generally, one can do this not only on the atoms of one context, but also on a selection of atoms
and joint probabilities of two or more contexts~\cite{pitowsky-89a,Pit-91,Pit-94,2000-poly}.
This results in what
Boole~\cite{Boole,Boole-62} called {\em ``conditions of possible experience''} for
the {\em ``concurrence of events.''}
In an Einstein-Podolsky-Rosen setup one ends up in Bell-type inequalities, which are
prominently violated by quantum probabilities and correlations.
Alas, the quantum correlations do not violate the inequalities maximally, which has led to
the introduction of so-called ``nonlocal boxes''~\cite{popescu-2014},
which may be obtained by ``sharpening''
the two-partite quantum correlations to a Heaviside function~\cite{svozil-krenn}.

As long as there are ``sufficiently many'' two-valued measures
(e.g. capable of separating two arbitrary atoms)
one might generalize this strategy to non-Boolean propositional structures.
In particular, one could obtain quasi-classical probability measures
by enumerating all two-valued measures, and by then taking the convex combination
(\ref{2015-s-e1}) thereof~\cite{svozil-2008-ql}.
One can do this because a two-valued measure has to ``cover'' all involved contexts simultaneously:
if subclassicality is assumed, then the same two-valued measure defined on one context contributes to
all the other contexts in such a way that the sums of that measure, taken along any such context has to
be additive and yield one.

\subsection{Cat's cradle configurations}

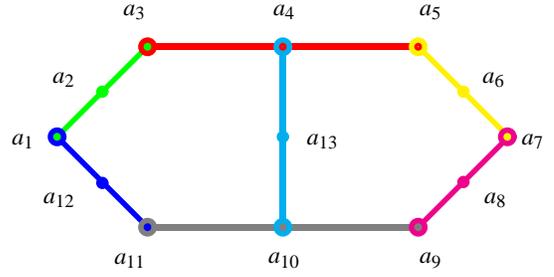
\begin{figure}
\begin{center}
\unitlength 0.6mm
\allinethickness{2.1pt}
\begin{picture}(108.00,55.00)
\put(25.00,7.33){\color{gray}\line(1,0){60.00}}
\put(25.00,47.33){\color{red}\line(1,0){60.00}}
\put(55.00,7.33){\color{cyan}\line(0,1){40.00}}
\put(25.00,7.33){\color{blue}\line(-1,1){20.00}}
\put(5.00,27.33){\color{green}\line(1,1){20.00}}
\put(85.00,7.33){\color{magenta}\line(1,1){20.00}}
\put(105.00,27.33){\color{yellow}\line(-1,1){20.00}}
\put(24.67,55.00){\makebox(0,0)[rc]{$a_3$}}
\put(55.33,55.00){\makebox(0,0)[cc]{$a_4$}}
\put(85.33,55.00){\makebox(0,0)[lc]{$a_5$}}
\put(9.00,40.00){\makebox(0,0)[rc]{$a_2$}}
\put(99.33,40.00){\makebox(0,0)[lc]{$a_6$}}
\put(0.00,26.33){\makebox(0,0)[rc]{$a_1$}}
\put(108.00,26.33){\makebox(0,0)[lc]{$a_7$}}
\put(60.33,26.33){\makebox(0,0)[lc]{$a_{13}$}}
\put(9.00,13.33){\makebox(0,0)[rc]{$a_{12}$}}
\put(99.67,13.33){\makebox(0,0)[lc]{$a_8$}}
\put(24.67,-0.05){\makebox(0,0)[rc]{$a_{11}$}}
\put(55.33,-0.05){\makebox(0,0)[cc]{$a_{10}$}}
\put(85.33,-0.05){\makebox(0,0)[lc]{$a_9$}}
\put(15.00,17.09){\color{blue}\circle{1.5}}
\put(25.00,7.33){\color{blue}\circle{1.5}}
\put(25.00,7.33){\color{gray}\circle{3.00}}
\put(55.00,27.33){\color{cyan}\circle{1.5}}
\put(85.00,7.33){\color{gray}\circle{1.5}}
\put(85.00,7.33){\color{magenta}\circle{3.00}}
\put(95.00,17.33){\color{magenta}\circle{1.5}}
\put(5.00,27.33){\color{green}\circle{1.5}}
\put(5.00,27.33){\color{blue}\circle{3.0}}
\put(15.00,37.33){\color{green}\circle{1.5}}
\put(25.00,47.33){\color{green}\circle{1.5}}
\put(25.00,47.33){\color{red}\circle{3.00}}
\put(55.00,47.33){\color{red}\circle{1.5}}
\put(55.00,47.33){\color{cyan}\circle{3.00}}
\put(85.00,47.33){\color{red}\circle{1.5}}
\put(85.00,47.33){\color{yellow}\circle{3.00}}
\put(55.00,7.33){\color{gray}\circle{1.5}}
\put(55.00,7.33){\color{cyan}\circle{3.00}}
\put(104.76,27.33){\color{yellow}\circle{1.5}}
\put(104.76,27.33){\color{magenta}\circle{3.00}}
\put(95.00,37.33){\color{yellow}\circle{1.5}}
\end{picture}
\end{center}
\caption{\label{2015-cesena-f2} (Color online) Orthogonality diagram of a cats cradle logic which requires that, for two-valued measures,
if $v(a_1)=1$, then $v(a_7)=0$. For a partition logic as well as for a Hilbert space realization
see Refs.~\cite{svozil-tkadlec,svozil-2008-ql}.}
\end{figure}

Consider a propositional structure depicted in Fig.~\ref{2015-cesena-f2}.
As Pitowsky~\cite{Pitowsky2003395,pitowsky-06} has pointed out,
the reduction of some probabilities of atoms at intertwined contexts yields
\begin{equation}
p_1+p_7=\frac{3}{2}- \frac{1}{2}\left(p_{12}+p_{13}+p_2+p_6+p_8\right)\le \frac{3}{2},
\label{2015-s-e2}
\end{equation}
because all probabilities $p_i$ are non-negative.
Indeed, if one applies the standard quantum mechanical Born (trace) rule to a particular realization
enumerated in Fig.~4 of Ref.~\cite{svozil-tkadlec},
then, as
$a_1\equiv \frac{1}{\sqrt{3}}\left( \sqrt{2},-1,0 \right)$
and
$a_7\equiv \frac{1}{\sqrt{3}}\left( \sqrt{2},1,0 \right)$,
the quantum probability of finding the quantum in a state spanned by $a_7$ if it has been prepared in a state
spanned by $a_1$ is
$p_7(a_1)= \langle a_7 \vert a_1 \rangle^2 = \frac{1}{9}$.
Together with  $p_1(a_1)= \langle a_1 \vert a_1 \rangle^2 = 1$
we obtain   $p_1(a_1) + p_7(a_1) = \frac{10}{9}$, which satisfies the classical bound $\frac{3}{2}$.

Indeed, a closer look at the quantum probabilities reveals that, with
$a_{13}\equiv  \left( 0, 1,0 \right)$,
$a_{6,8}\equiv \frac{1}{2\sqrt{3}}\left( -1,\sqrt{2},\pm 3 \right)$,
$p_{12}(a_1) =p_2(a_1)=0$,
$p_{13}(a_1)= \frac{1}{3}$, and
$p_6(a_1)=p_8(a_1) = \frac{4}{9}$,
the classical bounds of probability~(\ref{2015-s-e2}) -- Boole's conditions of possible experience --
are perfectly satisfied by the quantum predictions, since
$
1+\frac{1}{9} =
\frac{3}{2} - \frac{1}{2}\left(0+\frac{1}{3}+0+\frac{2}{9}+\frac{2}{9}\right)$.
This was to be expected, as Eq.~(\ref{2015-s-e2}) has been derived by supposing subclassicality
which is satisfied both by quasi-classical (e.g. generalized urn as well as automata)
models as well as quantum mechanics.

\begin{figure}
\begin{center}
\begin{tabular}{ccc}
\unitlength 0.3mm
\allinethickness{1.5pt}
\begin{picture}(108.00,55.00)
\put(25.00,7.33){\color{gray}\line(1,0){60.00}}
\put(25.00,47.33){\color{red}\line(1,0){60.00}}
\put(55.00,7.33){\color{cyan}\line(0,1){40.00}}
\put(25.00,7.33){\color{blue}\line(-1,1){20.00}}
\put(5.00,27.33){\color{green}\line(1,1){20.00}}
\put(85.00,7.33){\color{magenta}\line(1,1){20.00}}
\put(105.00,27.33){\color{yellow}\line(-1,1){20.00}}
\put(24.67,55.00){\makebox(0,0)[rc]{$0$}}
\put(55.33,58.00){\makebox(0,0)[cc]{$\frac{1}{3}$}}
\put(87,58.00){\makebox(0,0)[lc]{$\frac{2}{3}$}}
\put(9.00,40.00){\makebox(0,0)[rc]{$0$}}
\put(99.33,40.00){\makebox(0,0)[lc]{$\frac{2}{9}$}}
\put(0.00,26.33){\makebox(0,0)[rc]{$1$}}
\put(110.00,26.33){\makebox(0,0)[lc]{$\frac{1}{9}$}}
\put(60.33,26.33){\makebox(0,0)[lc]{$\frac{1}{3}$}}
\put(9.00,13.33){\makebox(0,0)[rc]{$0$}}
\put(99.67,13.33){\makebox(0,0)[lc]{$\frac{2}{9}$}}
\put(24.67,-2){\makebox(0,0)[rc]{$0$}}
\put(55.33,-3){\makebox(0,0)[cc]{$\frac{1}{3}$}}
\put(87,-2){\makebox(0,0)[lc]{$\frac{2}{3}$}}
\put(15.00,17.09){\color{blue}\circle{1.5}}
\put(25.00,7.33){\color{blue}\circle{1.5}}
\put(25.00,7.33){\color{gray}\circle{3.00}}
\put(55.00,27.33){\color{cyan}\circle{1.5}}
\put(85.00,7.33){\color{gray}\circle{1.5}}
\put(85.00,7.33){\color{magenta}\circle{3.00}}
\put(95.00,17.33){\color{magenta}\circle{1.5}}
\put(5.00,27.33){\color{green}\circle{1.5}}
\put(5.00,27.33){\color{blue}\circle{3.0}}
\put(15.00,37.33){\color{green}\circle{1.5}}
\put(25.00,47.33){\color{green}\circle{1.5}}
\put(25.00,47.33){\color{red}\circle{3.00}}
\put(55.00,47.33){\color{red}\circle{1.5}}
\put(55.00,47.33){\color{cyan}\circle{3.00}}
\put(85.00,47.33){\color{red}\circle{1.5}}
\put(85.00,47.33){\color{yellow}\circle{3.00}}
\put(55.00,7.33){\color{gray}\circle{1.5}}
\put(55.00,7.33){\color{cyan}\circle{3.00}}
\put(104.76,27.33){\color{yellow}\circle{1.5}}
\put(104.76,27.33){\color{magenta}\circle{3.00}}
\put(95.00,37.33){\color{yellow}\circle{1.5}}
\end{picture}
&&
\unitlength 0.3mm
\allinethickness{1.5pt}
\begin{picture}(108.00,55.00)
\put(25.00,7.33){\color{gray}\line(1,0){60.00}}
\put(25.00,47.33){\color{red}\line(1,0){60.00}}
\put(55.00,7.33){\color{cyan}\line(0,1){40.00}}
\put(25.00,7.33){\color{blue}\line(-1,1){20.00}}
\put(5.00,27.33){\color{green}\line(1,1){20.00}}
\put(85.00,7.33){\color{magenta}\line(1,1){20.00}}
\put(105.00,27.33){\color{yellow}\line(-1,1){20.00}}
\put(24.67,55.00){\makebox(0,0)[rc]{$0$}}
\put(55.33,55.00){\makebox(0,0)[cc]{$y$}}
\put(85.33,55.00){\makebox(0,0)[lc]{$x+z$}}
\put(9.00,40.00){\makebox(0,0)[rc]{$0$}}
\put(99.33,40.00){\makebox(0,0)[lc]{$y$}}
\put(0.00,26.33){\makebox(0,0)[rc]{$1$}}
\put(110.00,26.33){\makebox(0,0)[lc]{$0$}}
\put(60.33,26.33){\makebox(0,0)[lc]{$x$}}
\put(9.00,13.33){\makebox(0,0)[rc]{$0$}}
\put(99.67,13.33){\makebox(0,0)[lc]{$z$}}
\put(24.67,-0.05){\makebox(0,0)[rc]{$0$}}
\put(55.33,-0.05){\makebox(0,0)[cc]{$z$}}
\put(85.33,-0.05){\makebox(0,0)[lc]{$x+y$}}
\put(15.00,17.09){\color{blue}\circle{1.5}}
\put(25.00,7.33){\color{blue}\circle{1.5}}
\put(25.00,7.33){\color{gray}\circle{3.00}}
\put(55.00,27.33){\color{cyan}\circle{1.5}}
\put(85.00,7.33){\color{gray}\circle{1.5}}
\put(85.00,7.33){\color{magenta}\circle{3.00}}
\put(95.00,17.33){\color{magenta}\circle{1.5}}
\put(5.00,27.33){\color{green}\circle{1.5}}
\put(5.00,27.33){\color{blue}\circle{3.0}}
\put(15.00,37.33){\color{green}\circle{1.5}}
\put(25.00,47.33){\color{green}\circle{1.5}}
\put(25.00,47.33){\color{red}\circle{3.00}}
\put(55.00,47.33){\color{red}\circle{1.5}}
\put(55.00,47.33){\color{cyan}\circle{3.00}}
\put(85.00,47.33){\color{red}\circle{1.5}}
\put(85.00,47.33){\color{yellow}\circle{3.00}}
\put(55.00,7.33){\color{gray}\circle{1.5}}
\put(55.00,7.33){\color{cyan}\circle{3.00}}
\put(104.76,27.33){\color{yellow}\circle{1.5}}
\put(104.76,27.33){\color{magenta}\circle{3.00}}
\put(95.00,37.33){\color{yellow}\circle{1.5}}
\end{picture}
\\
$\;$
\\
(i)&$\qquad$&(ii)
\\
\multicolumn{3}{c}{
\unitlength 0.4mm
\allinethickness{1.8pt}
\begin{picture}(108.00,80.00)(0,-10)
\put(25.00,7.33){\color{gray}\line(1,0){60.00}}
\put(25.00,47.33){\color{red}\line(1,0){60.00}}
\put(55.00,7.33){\color{cyan}\line(0,1){40.00}}
\put(25.00,7.33){\color{blue}\line(-1,1){20.00}}
\put(5.00,27.33){\color{green}\line(1,1){20.00}}
\put(85.00,7.33){\color{magenta}\line(1,1){20.00}}
\put(105.00,27.33){\color{yellow}\line(-1,1){20.00}}
\put(24.67,61.00){\makebox(0,0)[rc]{\scriptsize $\lambda_{10} +\lambda_{11}+$}}
\put(24.67,55.00){\makebox(0,0)[rc]{\scriptsize $+ \lambda_{12} + \lambda_{13} + \lambda_{14}$}}
\put(55.33,61.00){\makebox(0,0)[cc]{\scriptsize $\lambda_2 + \lambda_6 + $}}
\put(55.33,55.00){\makebox(0,0)[cc]{\scriptsize $+ \lambda_7 + \lambda_8$}}
\put(85.33,61.00){\makebox(0,0)[lc]{\scriptsize $\lambda_1 + \lambda_3 + \lambda_4 +$}}
\put(85.33,55.00){\makebox(0,0)[lc]{\scriptsize $+ \lambda_{12} + \lambda_{13} + \lambda_{14}$}}
\put(9.00,43.00){\makebox(0,0)[rc] {\scriptsize $\lambda_4 + \lambda_5 + \lambda_6 + $}}
\put(9.00,37.00){\makebox(0,0)[rc] {\scriptsize $ + \lambda_7 + \lambda_8 + \lambda_9$}}
\put(99.33,43.00){\makebox(0,0)[lc]{\scriptsize $\lambda_2 + \lambda_6 + \lambda_8 +$}}
\put(99.33,37.00){\makebox(0,0)[lc]{\scriptsize $ + \lambda_{11} + \lambda_{12} + \lambda_{14}$}}
\put(0.00,26.33){\makebox(0,0)[rc] {\scriptsize $\lambda_1 + \lambda_2 + \lambda_3$}}
\put(108.00,26.33){\makebox(0,0)[lc]{\scriptsize $\lambda_7 + \lambda_{10} + \lambda_{13}$}}
\put(60.33,32.33){\makebox(0,0)[lc]{\scriptsize $\lambda_1 + \lambda_4 + \lambda_5 + $}}
\put(60.33,26.33){\makebox(0,0)[lc]{\scriptsize $+ \lambda_{10} + \lambda_{11} + $}}
\put(60.33,20.33){\makebox(0,0)[lc]{\scriptsize $+ \lambda_{12}$}}
\put(9.00,15.33){\makebox(0,0)[rc] {\scriptsize $\lambda_4 + \lambda_6 + \lambda_9 + $}}
\put(9.00,9.33){\makebox(0,0)[rc]  {\scriptsize $+ \lambda_{12} + \lambda_{13} + \lambda_{14}$}}
\put(99.67,15.33){\makebox(0,0)[lc]{\scriptsize $\lambda_3 + \lambda_5 + \lambda_8 + $}}
\put(99.67,9.33){\makebox(0,0)[lc] {\scriptsize $+ \lambda_9 + \lambda_{11} + \lambda_{14}$}}
\put(24.67,-0.05){\makebox(0,0)[rc]{\scriptsize $\lambda_5 + \lambda_7 + \lambda_8 +$}}
\put(24.67,-6.05){\makebox(0,0)[rc]{\scriptsize $+ \lambda_{10} + \lambda_{11}$}}
\put(55.33,-0.05){\makebox(0,0)[cc]{\scriptsize $\lambda_3 + \lambda_9 + $}}
\put(55.33,-6.05){\makebox(0,0)[cc]{\scriptsize $ + \lambda_{13} + \lambda_{14}$}}
\put(85.33,-0.05){\makebox(0,0)[lc]{\scriptsize $\lambda_1 + \lambda_2 + \lambda_4 + $}}
\put(85.33,-6.05){\makebox(0,0)[lc]{\scriptsize $ + \lambda_6 + \lambda_{12}$}}
\put(15.00,17.09){\color{blue}\circle{1.5}}
\put(25.00,7.33){\color{blue}\circle{1.5}}
\put(25.00,7.33){\color{gray}\circle{3.00}}
\put(55.00,27.33){\color{cyan}\circle{1.5}}
\put(85.00,7.33){\color{gray}\circle{1.5}}
\put(85.00,7.33){\color{magenta}\circle{3.00}}
\put(95.00,17.33){\color{magenta}\circle{1.5}}
\put(5.00,27.33){\color{green}\circle{1.5}}
\put(5.00,27.33){\color{blue}\circle{3.0}}
\put(15.00,37.33){\color{green}\circle{1.5}}
\put(25.00,47.33){\color{green}\circle{1.5}}
\put(25.00,47.33){\color{red}\circle{3.00}}
\put(55.00,47.33){\color{red}\circle{1.5}}
\put(55.00,47.33){\color{cyan}\circle{3.00}}
\put(85.00,47.33){\color{red}\circle{1.5}}
\put(85.00,47.33){\color{yellow}\circle{3.00}}
\put(55.00,7.33){\color{gray}\circle{1.5}}
\put(55.00,7.33){\color{cyan}\circle{3.00}}
\put(104.76,27.33){\color{yellow}\circle{1.5}}
\put(104.76,27.33){\color{magenta}\circle{3.00}}
\put(95.00,37.33){\color{yellow}\circle{1.5}}
\end{picture}
}
\\
\multicolumn{3}{c}{(iii)}
\end{tabular}
\end{center}
\caption{\label{2015-cesena-f3} (Color online) Orthogonality diagram of the logic depicted
in Fig.~\ref{2015-cesena-f2}
with overlaid (i) quantum and
(ii) classical prediction probabilities for a state prepared along $a_1$.
The classical predictions require that $x$, $y$ and $z$ are non-negative and $x+y+z = 1$.
(iii) The full parametrization of classical probabilities;
with non-negative $\lambda_1,\ldots \lambda_{14}\ge 0$, and $\lambda_1+\cdots +\lambda_{14} = 1$.
Note that the special case (ii) is obtained by identifying
with
$\lambda_1=x$,
$\lambda_2=y$,
$\lambda_3=z$,
and  $\lambda_4,\ldots \lambda_{14}= 0$.
}
\end{figure}
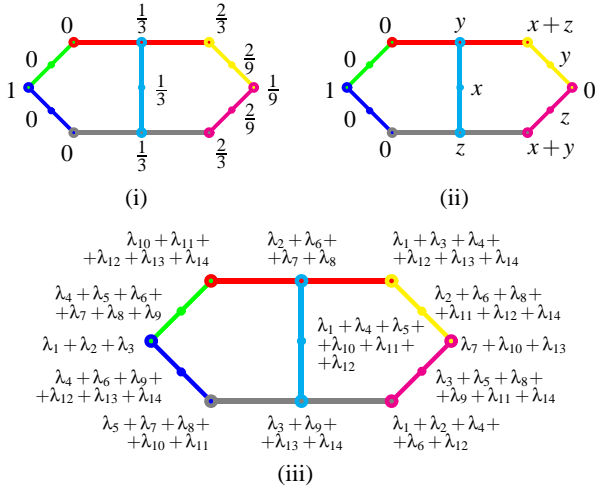

But does that mean that the classical and quantum predictions coincide?
The quantum predictions, computed under the assumption that the system is prepared in state $a_1$ and thus $p_1(a_1)=1$,
are enumerated in Fig.~\ref{2015-cesena-f3}(i).
Note that the sum of the probabilities of each context has to sum up to unity.

In contrast to the quantum predictions, with the same preparation,
the classical predictions cannot yield any $p_7(a_1)$ other than zero, because
by the way the logic is constructed there does not exist any two-valued measure satisfying
$p_1(a_1)=p_7(a_1)=1$.
(This is easily derivable by proving the impossibility of any such measure~\cite{svozil-2006-omni}.)
They are enumerated in Fig.~\ref{2015-cesena-f3}(ii).
The full parametrization of all conceivable classical probabilities is depicted in  Fig.~\ref{2015-cesena-f3}(iii).

So, if one interprets this argument in terms of a (state dependent) Boole-Bell type inequality,
all it needs is to prepare a three-state quantum system in a state along $a_1\equiv \frac{1}{\sqrt{3}}\left( \sqrt{2},-1,0 \right)$
and measure the projection observable along $a_7\equiv \frac{1}{\sqrt{3}}\left( \sqrt{2},1,0 \right)$.
In a generalized beam splitter setup~\cite{rzbb},
once the detector associated with $a_7$ clicks on the input associated with port $a_1$
one knows that the underlying physical realization is ``quantum-like'' and not classical.
This represents another type of violation of Boole's conditions of possible experience by quantized systems.

There exist more quantum predictions contradicting
(quasi-) classical predictions based on additivity:
suppose a tandem cat's cradle logic, which are just two cat's cradle logics intertwined at three contexts per copy,
with a non-separating set of two-valued states
already discussed by Kochen and Specker~\cite[$\Gamma_3$, p.~70]{kochen1},
and explicitly parameterized in three-dimensional
real Hilbert space by Tkadlec~\cite[Fig.~1]{tkadlec-96},
thereby continuing the observables and preparations already used earlier.
Classical predictions based on this set of observables
would require that
that if one prepares a quantized system in
$a_1\equiv \frac{1}{\sqrt{3}}\left( \sqrt{2},-1,0 \right)$
and measure it along
$b\equiv \frac{1}{\sqrt{3}}\left( -1,\sqrt{2},0 \right)$,
the measurement would always yield a positive result, because every two-valued measure
$v$ on that logic must satisfy $v(a_1)=v(b)=1$.
However, the quantum predictions,
also satisfying subclassicality, are $\langle b \vert a_1 \rangle^2 =\frac{8}{9}$.

The full hull computation~\cite{cdd-pck}
reveals the Boole-Bell type conditions of possible experience
\begin{equation}
\begin{split}
 p_1+p_2+p_6\geq p_4+p_8, \\
 p_1+p_2\geq p_4, \\
 p_1+2 p_2+p_6\geq 2 p_4+p_8, \\
 p_2+p_6\geq p_4,  \ldots \\
 p_{10}+p_2+p_6\geq p_4+p_8, \\
 p_4+p_8+1\geq p_1+p_{10}+p_2+p_6, \\
 p_8+1\geq p_1+p_{10}+p_2, \\
 p_4+1\geq p_1+p_2+p_6, \\
 p_4+p_5\geq p_1+p_2, \\
 p_1+p_2+p_6+p_7\geq p_4+1, \\
 p_4+p_8+p_9\geq p_1+p_2+p_6, \\
 p_1+p_{10}+p_{11}+p_2+p_6\geq p_4+p_8+1, \\
 p_{12}+p_4+p_8\geq p_{10}+p_2+p_6, \\
 p_{10}+p_{13}+p_4\geq 1
\end{split}
\label{2015-s-e-ccbb}
\end{equation}
as bounds of the polytope spanned by the two-valued measures interpreted as vertices.
Some of these classical bounds
are enumerated in Eq.~(\ref{2015-s-e-ccbb}).
A fraction of these, in particular, $p_2+p_6\geq p_4$
is violated by the quantum probabilities mentioned earlier,
as
$p_2=0$,
$p_6=\frac{2}{9}$,
and
$p_4=\frac{1}{3}$.

\subsection{Pentagon configuration}

\begin{figure}
\begin{center}
\begin{tabular}{ccc}
\unitlength 0.16mm
\allinethickness{2.1pt}
\begin{picture}(230,200)(-110,-100)
\multiput(31,-95.25)(.033724340176,.046554252199){2046}{\color{cyan}\line(0,1){.046554252199}}
\multiput(100,0)(-.033724340176,.046432062561){2046}{\color{magenta}\line(0,1){.046432062561}}
\multiput(31,95)(-.10418604651,-.03372093023){1075}{\color{blue}\line(-1,0){.10418604651}}
\put(-81,58.75){\color{red}\line(0,-1){117.75}}
\multiput(-81,-59)(.10328096118,-.03373382625){1082}{\color{green}\line(1,0){.10328096118}}
%
\put( 30.9017 , 95.1057){\color{blue}\circle{1.20}} 
\put( 30.9017 , 95.1057){\color{magenta}\circle{9.00}}
\put( 55.9017 , 95.1057){\makebox(0,0)[cc]{$a_1$}}
\put(100,0){\color{magenta}\circle{1.20}}    
\put(100,0){\color{cyan}\circle{9.00}}
\put(120,0){\makebox(0,0)[cc]{$a_3$}}
\put( 30.9017 , -95.1057){\color{cyan}\circle{1.20}}  
\put( 30.9017 , -95.1057){\color{green}\circle{9.00}}
\put( 55.9017 , -95.1057){\makebox(0,0)[cc]{$a_5$}}
\put( -80.9017 , -58.7785){\color{green}\circle{1.20}}   
\put( -80.9017 , -58.7785){\color{red}\circle{9.00}}
\put( -105.9017 , -58.7785){\makebox(0,0)[cc]{$a_7$}}
\put(-80.9017 , 58.7785){\color{red}\circle{1.20}}     
\put(-80.9017 , 58.7785){\color{blue}\circle{9.00}}
\put(-105.9017 , 58.7785){\makebox(0,0)[cc]{$a_9$}}
\end{picture}
&&
\unitlength 0.16mm
\allinethickness{2.1pt}
\begin{picture}(230,200)(-110,-100)
\multiput(31,-95.25)(.033724340176,.046554252199){2046}{\color{cyan}\line(0,1){.046554252199}}
\multiput(100,0)(-.033724340176,.046432062561){2046}{\color{magenta}\line(0,1){.046432062561}}
\multiput(31,95)(-.10418604651,-.03372093023){1075}{\color{blue}\line(-1,0){.10418604651}}
\put(-81,58.75){\color{red}\line(0,-1){117.75}}
\multiput(-81,-59)(.10328096118,-.03373382625){1082}{\color{green}\line(1,0){.10328096118}}
%
\put( 30.9017 , 95.1057){\color{blue}\circle{1.20}} 
\put( 30.9017 , 95.1057){\color{magenta}\circle{9.00}}
\put( 55.9017 , 95.1057){\makebox(0,0)[cc]{$a_1$}}
\put( 65.4509,47.5529){\color{magenta}\circle{9}}  
\put( 90.4509,47.5529){\makebox(0,0)[cc]{$a_2$}}
\put(100,0){\color{magenta}\circle{1.20}}    
\put(100,0){\color{cyan}\circle{9.00}}
\put(120,0){\makebox(0,0)[cc]{$a_3$}}
\put( 65.4509,-47.5529){\color{cyan}\circle{9}}  
\put( 90.4509,-47.5529){\makebox(0,0)[cc]{$a_4$}}
\put( 30.9017 , -95.1057){\color{cyan}\circle{1.20}}  
\put( 30.9017 , -95.1057){\color{green}\circle{9.00}}
\put( 55.9017 , -95.1057){\makebox(0,0)[cc]{$a_5$}}
\put( -25,-76.9421){\color{green}\circle{9}}         
\put( -40,-90.9421){\makebox(0,0)[cc]{$a_6$}}
\put( -80.9017 , -58.7785){\color{green}\circle{1.20}}   
\put( -80.9017 , -58.7785){\color{red}\circle{9.00}}
\put( -105.9017 , -58.7785){\makebox(0,0)[cc]{$a_7$}}
\put(-80.9017,0){\color{red}\circle{9}}           
\put(-105.9017,0){\makebox(0,0)[cc]{$a_8$}}
\put(-80.9017 , 58.7785){\color{red}\circle{1.20}}     
\put(-80.9017 , 58.7785){\color{blue}\circle{9.00}}
\put(-105.9017 , 58.7785){\makebox(0,0)[cc]{$a_9$}}
\put( -25,76.9421){\color{blue}\circle{9}}         
\put( -40,90.9421){\makebox(0,0)[cc]{$a_{10}$}}
\end{picture}
\\
$\;$
(i)&$\quad$&(ii)
\end{tabular}
\end{center}
\caption{\label{2015-s-f6} (Color online) Orthogonality diagram of the reduced pentagon (i),
and of the pentagon logic (ii).
A realization of (ii) in terms of partition logic is enumerated in Eq.~(\ref{2015-s-e6});
an explicit quantum realizaion can be found in Ref.~\cite{svozil-tkadlec}.}
\end{figure}
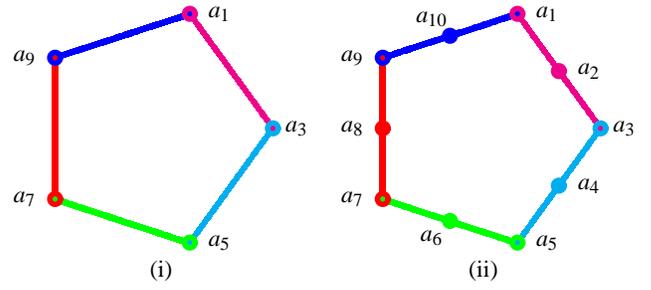

There exist, however, probabilities that are neither quasi-classical nor quantum-like although they satisfy subclassicality,
and although the underlying logic can be realized both quasi-classically by partition logics as well as quantum mechanically.
For the sake of an example, we shall discuss  Wright's dispersionless state~\cite{wright:pent} on the logic whose orthogonality diagram is a pentagon,
as depicted in Fig.~\ref{2015-s-f6}(ii).

What are the probabilities of prediction associated with such structures?
The propositional structure depicted in Fig.~\ref{2015-s-f6}(i)
has no two-valued state, and just allows a single probability measure
which is constant on all atoms;
that is, $p_1=p_3=p_5=p_7=p_9=\frac{1}{2}$.

\begin{figure}
\begin{center}
\begin{tabular}{ccc}
\unitlength 0.1mm
\allinethickness{1.5pt}
\begin{picture}(230,200)(-110,-100)
\put(0,0){\makebox(0,0)[cc]{\large $v_1$}}
\multiput(31,-95.25)(1.2,1.6565){58}{\color{cyan}\line(0,1){.1656521739}}
\multiput(100,0)(-1.2,1.6522){58}{\color{magenta}\line(0,1){.1652173913}}
\multiput(31,95)(-3.69637,-1.19637){30}{\color{blue}\line(-1,0){.3696369637}}
\multiput(-81,59)(0,-2){60}{\color{red}\line(0,-1){0.33}}
\multiput(-81,-59)(3.664,-1.1967){31}{\color{green}\line(1,0){.3663934426}}
%
\put( 30.9017 , 95.1057){\circle{8}} 
\put( 65.4509,-47.5529){\circle{8}}  
\put( -25,-76.9421){\circle{8}}         
\put(-80.9017,0){\circle{8}}           
\end{picture}
&
\unitlength 0.1mm
\allinethickness{1.5pt}
\begin{picture}(230,200)(-110,-100)
\put(0,0){\makebox(0,0)[cc]{\large $v_2$}}
\multiput(31,-95.25)(1.2,1.6565){58}{\color{cyan}\line(0,1){.1656521739}}
\multiput(100,0)(-1.2,1.6522){58}{\color{magenta}\line(0,1){.1652173913}}
\multiput(31,95)(-3.69637,-1.19637){30}{\color{blue}\line(-1,0){.3696369637}}
\multiput(-81,59)(0,-2){60}{\color{red}\line(0,-1){0.33}}
\multiput(-81,-59)(3.664,-1.1967){31}{\color{green}\line(1,0){.3663934426}}
%
\put( 30.9017 , 95.1057){\circle{8}} 
\put( 30.9017 , -95.1057){\circle{8}}  
\put(-80.9017,0){\circle{8}}           
\end{picture}
&
\unitlength 0.1mm
\allinethickness{1.5pt}
\begin{picture}(230,200)(-110,-100)
\put(0,0){\makebox(0,0)[cc]{\large $v_3$}}
\multiput(31,-95.25)(1.2,1.6565){58}{\color{cyan}\line(0,1){.1656521739}}
\multiput(100,0)(-1.2,1.6522){58}{\color{magenta}\line(0,1){.1652173913}}
\multiput(31,95)(-3.69637,-1.19637){30}{\color{blue}\line(-1,0){.3696369637}}
\multiput(-81,59)(0,-2){60}{\color{red}\line(0,-1){0.33}}
\multiput(-81,-59)(3.664,-1.1967){31}{\color{green}\line(1,0){.3663934426}}
%
\put( 30.9017 , 95.1057){\circle{8}} 
\put( 65.4509,-47.5529){\circle{8}}  
\put( -80.9017 , -58.7785){\circle{8}}   
\end{picture}
\\
\unitlength 0.1mm
\allinethickness{1.5pt}
\begin{picture}(230,200)(-110,-100)
\put(0,0){\makebox(0,0)[cc]{\large $v_4$}}
\multiput(31,-95.25)(1.2,1.6565){58}{\color{cyan}\line(0,1){.1656521739}}
\multiput(100,0)(-1.2,1.6522){58}{\color{magenta}\line(0,1){.1652173913}}
\multiput(31,95)(-3.69637,-1.19637){30}{\color{blue}\line(-1,0){.3696369637}}
\multiput(-81,59)(0,-2){60}{\color{red}\line(0,-1){0.33}}
\multiput(-81,-59)(3.664,-1.1967){31}{\color{green}\line(1,0){.3663934426}}
%
\put(100,0){\circle{8}}    
\put( -25,-76.9421){\circle{8}}         
\put(-80.9017,0){\circle{8}}           
\put( -25,76.9421){\circle{8}}         
\end{picture}
&
\unitlength 0.1mm
\allinethickness{1.5pt}
\begin{picture}(230,200)(-110,-100)
\put(0,0){\makebox(0,0)[cc]{\large $v_5$}}
\multiput(31,-95.25)(1.2,1.6565){58}{\color{cyan}\line(0,1){.1656521739}}
\multiput(100,0)(-1.2,1.6522){58}{\color{magenta}\line(0,1){.1652173913}}
\multiput(31,95)(-3.69637,-1.19637){30}{\color{blue}\line(-1,0){.3696369637}}
\multiput(-81,59)(0,-2){60}{\color{red}\line(0,-1){0.33}}
\multiput(-81,-59)(3.664,-1.1967){31}{\color{green}\line(1,0){.3663934426}}
%
\put(100,0){\circle{8}}    
\put( -80.9017 , -58.7785){\circle{8}}   
\put( -25,76.9421){\circle{8}}         
\end{picture}
&
\unitlength 0.1mm
\allinethickness{1.5pt}
\begin{picture}(230,200)(-110,-100)
\put(0,0){\makebox(0,0)[cc]{\large $v_6$}}
\multiput(31,-95.25)(1.2,1.6565){58}{\color{cyan}\line(0,1){.1656521739}}
\multiput(100,0)(-1.2,1.6522){58}{\color{magenta}\line(0,1){.1652173913}}
\multiput(31,95)(-3.69637,-1.19637){30}{\color{blue}\line(-1,0){.3696369637}}
\multiput(-81,59)(0,-2){60}{\color{red}\line(0,-1){0.33}}
\multiput(-81,-59)(3.664,-1.1967){31}{\color{green}\line(1,0){.3663934426}}
%
\put(100,0){\circle{8}}    
\put( -25,-76.9421){\circle{8}}         
\put(-80.9017 , 58.7785){\circle{8}}     
\end{picture}
\\
\unitlength 0.1mm
\allinethickness{1.5pt}
\begin{picture}(230,200)(-110,-100)
\put(0,0){\makebox(0,0)[cc]{\large $v_7$}}
\multiput(31,-95.25)(1.2,1.6565){58}{\color{cyan}\line(0,1){.1656521739}}
\multiput(100,0)(-1.2,1.6522){58}{\color{magenta}\line(0,1){.1652173913}}
\multiput(31,95)(-3.69637,-1.19637){30}{\color{blue}\line(-1,0){.3696369637}}
\multiput(-81,59)(0,-2){60}{\color{red}\line(0,-1){0.33}}
\multiput(-81,-59)(3.664,-1.1967){31}{\color{green}\line(1,0){.3663934426}}
%
\put( 65.4509,47.5529){\circle{8}}  
\put( 30.9017 , -95.1057){\circle{8}}  
\put(-80.9017,0){\circle{8}}           
\put( -25,76.9421){\circle{8}}         
\end{picture}
&
\unitlength 0.1mm
\allinethickness{1.5pt}
\begin{picture}(230,200)(-110,-100)
\put(0,0){\makebox(0,0)[cc]{\large $v_8$}}
\multiput(31,-95.25)(1.2,1.6565){58}{\color{cyan}\line(0,1){.1656521739}}
\multiput(100,0)(-1.2,1.6522){58}{\color{magenta}\line(0,1){.1652173913}}
\multiput(31,95)(-3.69637,-1.19637){30}{\color{blue}\line(-1,0){.3696369637}}
\multiput(-81,59)(0,-2){60}{\color{red}\line(0,-1){0.33}}
\multiput(-81,-59)(3.664,-1.1967){31}{\color{green}\line(1,0){.3663934426}}
%
\put( 65.4509,47.5529){\circle{8}}  
\put( 30.9017 , -95.1057){\circle{8}}  
\put(-80.9017 , 58.7785){\circle{8}}     
\end{picture}
&
\unitlength 0.1mm
\allinethickness{1.5pt}
\begin{picture}(230,200)(-110,-100)
\put(0,0){\makebox(0,0)[cc]{\large $v_9$}}
\multiput(31,-95.25)(1.2,1.6565){58}{\color{cyan}\line(0,1){.1656521739}}
\multiput(100,0)(-1.2,1.6522){58}{\color{magenta}\line(0,1){.1652173913}}
\multiput(31,95)(-3.69637,-1.19637){30}{\color{blue}\line(-1,0){.3696369637}}
\multiput(-81,59)(0,-2){60}{\color{red}\line(0,-1){0.33}}
\multiput(-81,-59)(3.664,-1.1967){31}{\color{green}\line(1,0){.3663934426}}
%
\put( 65.4509,47.5529){\circle{8}}  
\put( 65.4509,-47.5529){\circle{8}}  
\put( -80.9017 , -58.7785){\circle{8}}   
\put( -25,76.9421){\circle{8}}         
\end{picture}
\\
\unitlength 0.1mm
\allinethickness{1.5pt}
\begin{picture}(230,200)(-110,-100)
\put(0,0){\makebox(0,0)[cc]{\large $v_{10}$}}
\multiput(31,-95.25)(1.2,1.6565){58}{\color{cyan}\line(0,1){.1656521739}}
\multiput(100,0)(-1.2,1.6522){58}{\color{magenta}\line(0,1){.1652173913}}
\multiput(31,95)(-3.69637,-1.19637){30}{\color{blue}\line(-1,0){.3696369637}}
\multiput(-81,59)(0,-2){60}{\color{red}\line(0,-1){0.33}}
\multiput(-81,-59)(3.664,-1.1967){31}{\color{green}\line(1,0){.3663934426}}
%
\put( 65.4509,47.5529){\circle{8}}  
\put( 65.4509,-47.5529){\circle{8}}  
\put( -25,-76.9421){\circle{8}}         
\put(-80.9017 , 58.7785){\circle{8}}     
\end{picture}
&
\unitlength 0.1mm
\allinethickness{1.5pt}
\begin{picture}(230,200)(-110,-100)
\put(0,0){\makebox(0,0)[cc]{\large $v_{11}$}}
\multiput(31,-95.25)(1.2,1.6565){58}{\color{cyan}\line(0,1){.1656521739}}
\multiput(100,0)(-1.2,1.6522){58}{\color{magenta}\line(0,1){.1652173913}}
\multiput(31,95)(-3.69637,-1.19637){30}{\color{blue}\line(-1,0){.3696369637}}
\multiput(-81,59)(0,-2){60}{\color{red}\line(0,-1){0.33}}
\multiput(-81,-59)(3.664,-1.1967){31}{\color{green}\line(1,0){.3663934426}}
%
\put( 65.4509,47.5529){\circle{8}}  
\put( 65.4509,-47.5529){\circle{8}}  
\put( -25,-76.9421){\circle{8}}         
\put(-80.9017,0){\circle{8}}           
\put( -25,76.9421){\circle{8}}         
\end{picture}
\end{tabular}
\end{center}
\caption{\label{2015-s-f7} Two-valued measures on the pentagon logic of Fig.~\ref{2015-s-f6}.}
\end{figure}
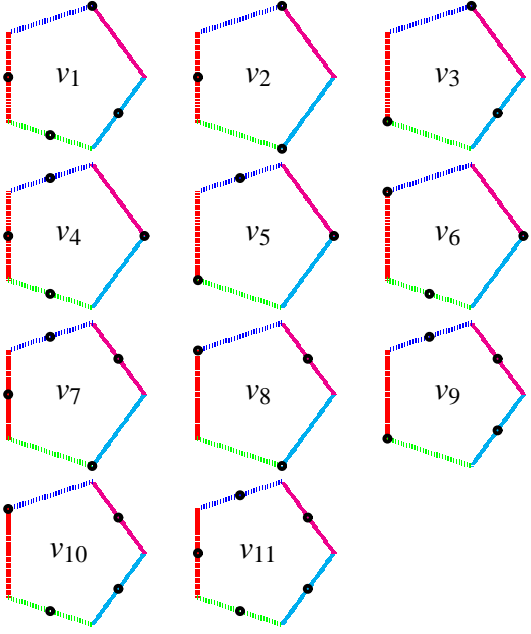

This prediction or oracle is still allowed by the subclassicality rule even if one adds one atom per block.
But, as has been pointed out by Wright~\cite{wright:pent}, it can neither be operationally realized
by any quasi-classical nor by any quantum oracle.
For quasi-classical systems, this can explicitly be demonstrated by enumerating
all two-valued measures on this ``pentagon logic'' of Fig.~\ref{2015-s-f6}(ii),
as depicted in Fig.~\ref{2015-s-f7}.
Note that no measure exists which is non-zero only on the atoms located at intertwining contexts; that is, which
does not vanish at one (or more) atoms at intertwining contexts,
and at the same time vanishes at all the ``middle'' atoms belonging to only one context.
Because the quasi-classical probabilities are just the convex sum Eq.~(\ref{2015-s-e1})
over all the two-valued measures it is clear that no classical probability vanishes at all non-intertwining atoms;
in particular one which is $\frac{1}{2}$ on all intertwining atoms.

A straightforward extraction~\cite{svozil-2001-eua,svozil-2008-ql}
based on two-valued measures in Fig.~\ref{2015-s-f7}
yields the partition logic
-- which is the pasting of subalgebras specified by partitions of the set $\{1,2, \ldots , 11\}$
in such a way that any atom is represented by the set of indices of two-valued measures
acquiring the value one on that atom
-- of indices of the two-valued measures
enumerated in Eq.~(\ref{2015-s-e6});
that is, in terms of the subscripts of the two-valued measures
(i.e., $v_i \rightarrow i$),
\begin{equation}
\begin{split}
\{
\{
\{ 1,2,3
\},
\{ 7,8,9,10,11
\},
\{ 4,5,6
\}
\}, \\
\{
\{ 4,5,6
\},
\{ 1,3,9,10,11
\},
\{ 2,7,8
\}
\}, \\
\{
\{ 2,7,8
\},
\{ 1,4,6,10,11
\},
\{ 3,5,9,3
\}
\},\\
\{
\{ 3,5,9,3
\},
\{ 1,2,4,7,11
\},
\{ 6,8,10
\}
\},\\
\{
\{ 6,8,10
\},
\{ 4,5,7,9,11
\},
\{ 1,2,3
\}
\}
\}
\end{split}
\label{2015-s-e6}
\end{equation}

\begin{figure}
\begin{center}
\begin{tabular}{c}
\unitlength 0.12mm
\allinethickness{2.1pt}
\begin{picture}(230,250)(-110,-115)
\multiput(31,-95.25)(.033724340176,.046554252199){2046}{\color{cyan}\line(0,1){.046554252199}}
\multiput(100,0)(-.033724340176,.046432062561){2046}{\color{magenta}\line(0,1){.046432062561}}
\multiput(31,95)(-.10418604651,-.03372093023){1075}{\color{blue}\line(-1,0){.10418604651}}
\put(-81,58.75){\color{red}\line(0,-1){117.75}}
\multiput(-81,-59)(.10328096118,-.03373382625){1082}{\color{green}\line(1,0){.10328096118}}
%
\put( 30.9017 , 95.1057){\color{blue}\circle{1.20}} 
\put( 30.9017 , 95.1057){\color{magenta}\circle{15.00}}
\put( 55.9017 , 95.1057){\makebox(0,0)[lc]{$\lambda_1 + \lambda_2 + \lambda_3$}}
\put( 65.4509,47.5529){\color{magenta}\circle{15.00}}  
\put( 90.4509,47.5529){\makebox(0,0)[lc]{$\lambda_7 + \lambda_8 + \lambda_9 + \lambda_{10} + \lambda_{11}$}}
\put(100,0){\color{magenta}\circle{1.20}}    
\put(100,0){\color{cyan}\circle{15.00}}
\put(120,0){\makebox(0,0)[lc]{$\lambda_4 + \lambda_5 + \lambda_6$}}
\put( 65.4509,-47.5529){\color{cyan}\circle{15.00}}  
\put( 90.4509,-47.5529){\makebox(0,0)[lc]{$\lambda_1 + \lambda_3 + \lambda_9 + \lambda_{10} + \lambda_{11}$}}
\put( 30.9017 , -95.1057){\color{cyan}\circle{1.20}}  
\put( 30.9017 , -95.1057){\color{green}\circle{15.00}}
\put(55.9017 , -95.1057){\makebox(0,0)[lc]{$\lambda_2 + \lambda_7 + \lambda_8$}}
\put( -25,-76.9421){\color{green}\circle{15.00}}         
\put( -40,-90.9421){\makebox(0,0)[rc]{$\lambda_1 + \lambda_4 + \lambda_6 + \lambda_{10} + \lambda_{11}$}}
\put( -80.9017 , -58.7785){\color{green}\circle{1.20}}   
\put( -80.9017 , -58.7785){\color{red}\circle{15.00}}
\put( -105.9017 , -58.7785){\makebox(0,0)[rc]{$\lambda_3 + \lambda_5 + \lambda_9 + \lambda_3$}}
\put(-80.9017,0){\color{red}\circle{15.00}}           
\put(-105.9017,0){\makebox(0,0)[rc]{$\lambda_1 + \lambda_2 + \lambda_4 + \lambda_7 + \lambda_{11}$}}
\put(-80.9017 , 58.7785){\color{red}\circle{1.20}}     
\put(-80.9017 , 58.7785){\color{blue}\circle{15.00}}
\put(-105.9017 , 58.7785){\makebox(0,0)[rc]{$\lambda_6 + \lambda_8 + \lambda_{10}$}}
\put( -25,76.9421){\color{blue}\circle{15.00}}         
\put( -40,90.9421){\makebox(0,0)[rc]{$\lambda_4 + \lambda_5 + \lambda_7 + \lambda_9 + \lambda_{11}$}}
\end{picture}
\end{tabular}
\end{center}
\caption{\label{2015-s-f8} (Color online) Classical probabilities on the pentagon logic.}
\end{figure}
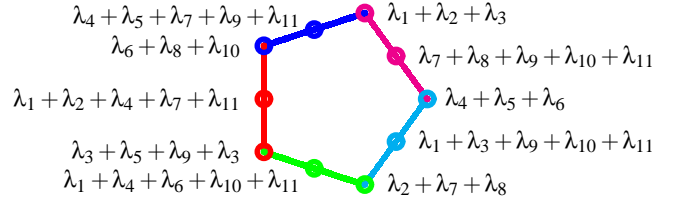

These partitions directly translate into the classical probabilities
which are, for instance, realizable by generalized urn or automaton models.
Fig.~\ref{2015-s-f8} parameterizes all classical probabilities through non-negative
$\lambda_1,\ldots ,\lambda_{11}\ge 0$ with
$\lambda_1+\cdots +\lambda_{11}=1$, subject to subclassicality.

The hull computation~\cite{cdd-pck}
reveals the Boole-Bell type conditions of possible experience
\begin{equation}
\begin{split}
p_4+p_8\geq p_1,  \ldots  \\
p_4+1\geq p_1+p_2+p_6,  \\
p_4+p_8+1\geq 2p_1+p_2+p_6,  \\
p_1+p_2\geq p_4,  \\
p_1+p_2+p_6\geq p_4+p_8,  \\
2p_1+p_{10}+p_2+p_6\geq p_4+p_8+1
\end{split}
\label{2015-s-e8}
\end{equation}
as bounds of the polytope spanned by the two-valued measures interpreted as vertices.
Some of these classical bounds are enumerated in Eq.~(\ref{2015-s-e8}).
Wright's measure, with $p_1=\frac{1}{2}$ and $p_4=p_8=0$, violates the first inequality.

\subsection{Triangle configurations}

\begin{figure}
\begin{center}
\begin{tabular}{ccccccc}
\unitlength 0.6mm 
\allinethickness{2.1pt}
\ifx\plotpoint\undefined\newsavebox{\plotpoint}\fi 
\begin{picture}(26,25)(-3,0)
\put(0,0){\color{blue}\line(1,0){20}}
\put(0,0){\color{red}\line(3,5){10}}
\put(20,0){\color{green}\line(-3,5){10}}
\put(0,0){\color{blue}\circle{1.2}}
\put(0,0){\color{red}\circle{3}}
\put(-2.5,0){\makebox(0,0)[rc]{$\frac{1}{2}$}}
\put(20,0){\color{green}\circle{1.2}}
\put(20,0){\color{blue}\circle{3}}
\put(22.5,0){\makebox(0,0)[lc]{$\frac{1}{2}$}}
\put(10,16.5){\color{red}\circle{1.2}}
\put(10,16.5){\color{green}\circle{3}}
\put(13,16.5){\makebox(0,0)[lc]{$\frac{1}{2}$}}
\end{picture}
&&
\unitlength 0.6mm 
\allinethickness{2.1pt}
\ifx\plotpoint\undefined\newsavebox{\plotpoint}\fi 
\begin{picture}(26,25)(-3,0)
\put(0,0){\color{blue}\line(1,0){20}}
\put(0,0){\color{red}\line(3,5){10}}
\put(20,0){\color{green}\line(-3,5){10}}
\put(0,0){\color{blue}\circle{1.2}}
\put(0,0){\color{red}\circle{3}}
\put(-3,0){\makebox(0,0)[rc]{$z$}}
\put(20,0){\color{green}\circle{1.2}}
\put(20,0){\color{blue}\circle{3}}
\put(23,0){\makebox(0,0)[lc]{$y$}}
\put(10,16.5){\color{red}\circle{1.2}}
\put(10,16.5){\color{green}\circle{3}}
\put(13,16.5){\makebox(0,0)[lc]{$x$}}
\put(5,8.25){\color{red}\circle{1.2}}
\put(3,9.25){\makebox(0,0)[rc]{$y$}}
\put(15,8.25){\color{green}\circle{1.2}}
\put(17,9.25){\makebox(0,0)[lc]{$z$}}
\put(10,0){\color{blue}\circle{1.2}}
\put(10,3){\makebox(0,0)[cc]{$x$}}
\end{picture}
&&
\unitlength 0.6mm 
\allinethickness{2.1pt}
\ifx\plotpoint\undefined\newsavebox{\plotpoint}\fi 
\begin{picture}(26,25)(-3,0)
\put(0,0){\color{blue}\line(1,0){20}}
\put(0,0){\color{red}\line(3,5){10}}
\put(20,0){\color{green}\line(-3,5){10}}
\put(0,0){\color{blue}\circle{1.2}}
\put(0,0){\color{red}\circle{3}}
\put(-2.5,0){\makebox(0,0)[rc]{$\frac{1}{2}$}}
\put(20,0){\color{green}\circle{1.2}}
\put(20,0){\color{blue}\circle{3}}
\put(22.5,0){\makebox(0,0)[lc]{$\frac{1}{2}$}}
\put(10,16.5){\color{red}\circle{1.2}}
\put(10,16.5){\color{green}\circle{3}}
\put(13,16.5){\makebox(0,0)[lc]{$\frac{1}{2}$}}
\put(5,8.25){\color{red}\circle{1.2}}
\put(3,9.25){\makebox(0,0)[rc]{$0$}}
\put(15,8.25){\color{green}\circle{1.2}}
\put(17,9.25){\makebox(0,0)[lc]{$0$}}
\put(10,0){\color{blue}\circle{1.2}}
\put(10,-5){\makebox(0,0)[cc]{$0$}}
\end{picture}
&&
\unitlength 0.6mm 
\allinethickness{2.1pt}
\ifx\plotpoint\undefined\newsavebox{\plotpoint}\fi 
\begin{picture}(26,25)(-3,0)
\put(0,0){\color{blue}\line(1,0){20}}
\put(0,0){\color{red}\line(3,5){10}}
\put(20,0){\color{green}\line(-3,5){10}}
\put(0,0){\color{blue}\circle{1.2}}
\put(0,0){\color{red}\circle{3}}
\put(-2.5,0){\makebox(0,0)[rc]{$\frac{1}{2}$}}
\put(20,0){\color{green}\circle{1.2}}
\put(20,0){\color{blue}\circle{3}}
\put(22.5,0){\makebox(0,0)[lc]{$\frac{1}{2}$}}
\put(10,16.5){\color{red}\circle{1.2}}
\put(10,16.5){\color{green}\circle{3}}
\put(13,16.5){\makebox(0,0)[lc]{$\frac{1}{2}$}}
\put(3.33,5.5){\color{red}\circle{1.2}}
\put(6.67,11){\color{red}\circle{1.2}}
\put(1,5.5){\makebox(0,0)[rc]{$0$}}
\put(4,11){\makebox(0,0)[rc]{$0$}}
\put(13.33,11){\color{green}\circle{1.2}}
\put(16.67,5.5){\color{green}\circle{1.2}}
\put(18.67,5.5){\makebox(0,0)[lc]{$0$}}
\put(15.33,11){\makebox(0,0)[lc]{$0$}}
\put(6.67,0){\color{blue}\circle{1.2}}
\put(13.33,0){\color{blue}\circle{1.2}}
\put(6.67,-5){\makebox(0,0)[cc]{$0$}}
\put(13.33,-5){\makebox(0,0)[cc]{$0$}}
\end{picture}
\\
$\;$\\
(i)&$\quad$&(ii)&$\quad$&(iii)&$\quad$&(iv)
\\
\multicolumn{7}{c}{
\unitlength 0.4mm 
\allinethickness{2.1pt}
\ifx\plotpoint\undefined\newsavebox{\plotpoint}\fi 
\begin{picture}(140,125)(0,0)
\put(20,20){\color{blue}\line(1,0){110}}
\multiput(20,20)(.03372164316,.05518087063){1631}{\color{red}\line(0,1){.05518087063}}
\multiput(75,110)(.03372164316,-.05518087063){1631}{\color{green}\line(0,-1){.05518087063}}
\put(20,20){\color{red}\circle{5}}
\put(20,20){\color{blue}\circle{2}}
\put(56.25,20){\color{blue}\circle{2}}
\put(56.25,20){\color{blue}\circle{5}}
\put(92.5,20){\color{blue}\circle{2}}
\put(92.5,20){\color{blue}\circle{5}}
\put(129.75,20){\color{green}\circle{2}}
\put(129.75,20){\color{blue}\circle{5}}
\put(56.25,79.75){\color{red}\circle{5}}
\put(56.25,79.75){\color{red}\circle{2}}
\put(38.75,51.25){\color{red}\circle{5}}
\put(38.75,51.25){\color{red}\circle{2}}
\put(74.75,109.75){\color{red}\circle{2}}
\put(74.75,109.75){\color{green}\circle{5}}
\put(93.75,79.75){\color{green}\circle{5}}
\put(93.75,79.75){\color{green}\circle{2}}
\put(111.25,51.25){\color{green}\circle{5}}
\put(111.25,51.25){\color{green}\circle{2}}
\put(15,11){\makebox(0,0)[rc]       {\footnotesize $p_1=\lambda_1 + \lambda_2$}}
\put(100,11){\makebox(0,0)[cc]     {\footnotesize $p_3=\lambda_8 + \lambda_9 + $}}
\put(100,3){\makebox(0,0)[cc]      {\footnotesize $ + \lambda_{10} + \lambda_{11} + \lambda_{12}$}}
\put(45,11){\makebox(0,0)[cc]       {\footnotesize $p_2=\lambda_3 + \lambda_4 +$}}
\put(45,3){\makebox(0,0)[cc]        {\footnotesize $  + \lambda_5+ \lambda_6 + \lambda_7$}}
\put(138,11){\makebox(0,0)[lc]      {\footnotesize $p_4=\lambda_{13} +\lambda_{14}$}}
\put(74.75,119){\makebox(0,0)[cc]  {\footnotesize $p_7=\lambda_3 + \lambda_8$}}
\put(108,80.25){\makebox(0,0)[lc]  {\footnotesize $p_6=\lambda_2 + \lambda_6 + \lambda_7 + $}}
\put(108,72.25){\makebox(0,0)[lc]  {\footnotesize $+ \lambda_{11} + \lambda_{12}$}}
\put(125,52.25){\makebox(0,0)[lc]  {\footnotesize $p_5=\lambda_1 + \lambda_4 + \lambda_5+$}}
\put(125,44.25){\makebox(0,0)[lc]  {\footnotesize $+ \lambda_9 + \lambda_{10}$}}
\put(42.5,80.25){\makebox(0,0)[rc] {\footnotesize $p_8=\lambda_4 + \lambda_6 + \lambda_9 +$}}
\put(42.5,72.25){\makebox(0,0)[rc] {\footnotesize $+ \lambda_{11} + \lambda_{13}$}}
\put(25.5,52.25){\makebox(0,0)[rc] {\footnotesize $p_9=\lambda_5 + \lambda_7 + \lambda_{10} +$}}
\put(25.5,44.25){\makebox(0,0)[rc] {\footnotesize $+ \lambda_{12} + \lambda_{14}$}}
\end{picture}
}
\\
\multicolumn{7}{c}{(v)}
\end{tabular}
\end{center}
\caption{(Color online) Classical probabilities (i) and (ii)
of the tight triangular pastings
of two- and three-atomic contexts introduced in Figs.~\ref{2015-s-f9}(i),(ii);
with
$x,y,z\ge 0$,
and
$x+y+z=1$.
The prediction probabilities represented by (iii) as well as (iv)
are neither classical nor quantum mechanical.
The classical probabilities on the triangle logic with four atoms per context
are enumerated in (v);
again $\lambda_1 , \ldots , \lambda_{14} \ge 0$ and
again $\lambda_1 + \cdots + \lambda_{14} = 1$.
\label{2015-s-f9}}
\end{figure}
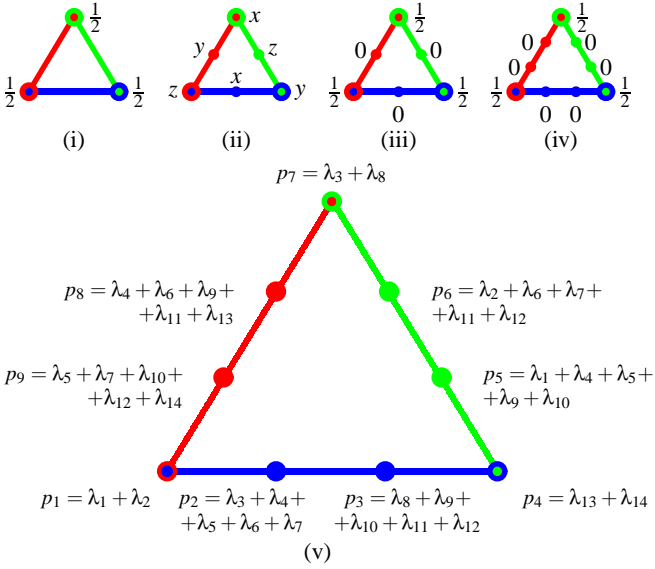

Very similar arguments hold also for the propositional structures
depicted in Figs.~\ref{2015-s-f2}(i),(ii):
Fig.~\ref{2015-s-f9}(i) represents a trivial classical prediction with equal probabilities.
Fig.~\ref{2015-s-f9}(ii) represents all classical predictions; the probability measures
being read off from the partition logic
$
\{
\{
\{1
\},
\{3
\},
\{2
\}
\},
\{
\{2
\},
\{1
\},
\{3
\}
\},
\{
\{3
\},
\{2
\},
\{1
\}
\}
\}
$
obtained from the three two-valued states on the logic in Fig.~\ref{2015-s-f2}(ii).
Figs.~\ref{2015-s-f9}(i),(iii) represent predictions $\frac{1}{2}$ for
all atoms at which the three contexts intertwine.
Fig.~\ref{2015-s-f9}(iii) represents a Wright prediction.
None of the propositional structures depicted in Figs.~\ref{2015-s-f9}(i)--(iii) allows a quantum realization.

Nevertheless,
in four-dimensional Hilbert space, the  propositional structure
with a triangular shaped orthogonality diagram allows a gemetric representation;
a particular one is explicitly enumerated in Fig.~4 of Ref.~\cite{2010-qchocolate}
whose classical probabilities are exhausted by the parameterization in Fig.~\ref{2015-s-f9}(v),
read off from the complete set of 14 two-valued measures enumerated in Fig.~5 of Ref.~\cite{2010-qchocolate}.
Fig.~\ref{2015-s-f9}(iv) represents a Wright prediction,
which cannot be realized classically as well as quantum mechanically for the same reasons as
mentioned earlier.
In the quantum case,
the proof of Theorem 2.2 of Ref.~\cite{wright:pent}
can be directly transferred to the four-dimensional configuration.

The hull computation~\cite{cdd-pck}
reveals the Boole-Bell type conditions of possible experience
\begin{equation}
\begin{split}
p_5+p_6\geq  p_1,  \ldots   \\
p_5+p_6+1\geq 2 p_1+p_2+p_3+p_8,   \\
p_1+p_2+p_3\geq p_5+p_6,   \\
p_5+p_6+p_7\geq p_1+p_2+p_3   \\
2p_1+p_2+p_3+p_8+p_9\geq p_5+p_6+1
\end{split}
\label{2015-s-e10}
\end{equation}
as bounds of the polytope spanned by the two-valued measures interpreted as vertices.
Some of these classical bounds are enumerated in Eq.~(\ref{2015-s-e10}).
Wright's measure, with $p_1=\frac{1}{2}$ and $p_5=p_6=0$, violates the first inequality.

\subsection{Gleason theorem and Kochen-Specker configurations}

The strategy to obtain predictions and probabilities by taking the convex sum of (sufficiently many)
two-valued measures satisfying subclassicality
fails completely for quantum systems with three or more mutually exclusive outcomes --
that is, for quantum Hilbert spaces of dimensions greater than two:
in this case, two-valued measures do not exist
even on certain finite substructures thereof~\cite{kochen1,2015-AnalyticKS}.

However, if one still clings to the subclassicality assumption
--
essentially requiring that
every context of maximally co-measurable observables is behaving
classically, and thus should be endowed with classical probabilities
--
then Gleason's theorem~\cite{Gleason,r:dvur-93,pitowsky:218,Peres-expTest-Glea}
derives the Born (trace) rule for quantum probabilities from subclassicality.
Indeed, as already observed by Gleason,
it is easy to see that, in the simplest case, such a subclassical (admissible) probability measure can be obtained in the form of a frame function $f_\rho$
by selecting some unit vector $\vert \rho \rangle$, corresponding to a pure quantum state (preparation),
and, for each closed subspace corresponding to a one-dimensional projection observable (i.e. an elementary yes-no proposition) $E=\vert e\rangle \langle e \vert$
along the unit vector $\vert e\rangle$,
and by taking
$f_\rho(\vert e\rangle ) = \langle \rho \vert e\rangle \langle e \vert  \rho \rangle = \vert \langle e \vert  \rho \rangle \vert^2$ as the square of the norm of the
projection of $\vert \rho \rangle$ onto the subspace spanned by $\vert e\rangle$.

The reason for this is that, because an arbitrary context can be represented as an orthonormal
basis $\{  \vert  e_i \rangle   \}$,
an {\it ad hoc}  frame function $f_\rho $
on any such context (and thus basis)
can be obtained by taking the length  of
the orthogonal (with respect to the basis vectors)
projections of $\vert  \rho \rangle$
onto all the basis vectors $\vert  e_i \rangle$,
that is, the norm of the resulting vector projections of $\vert \rho \rangle$ onto the basis vectors,
respectively.
This amounts to computing the absolute value of the Euclidean scalar products
$\langle  e_i \vert  \rho \rangle$
of the state vector with all the basis vectors.
In order that all such  absolute values of the scalar products (or the associated norms)
sum up to one and yield a frame function of weight one,
recall that $\vert  \rho \rangle$ is a unit vector
and note that, by the Pythagorean theorem,
these  absolute values of the individual scalar products
-- or the associated norms of the vector projections of $\vert \rho \rangle$ onto the basis vectors --
must be squared.
Thus the value $f_\rho(\vert e_i\rangle )$ of the frame function on the argument $\vert e_i\rangle $
must be the square of the scalar product of $\vert  \rho \rangle$
with $\vert  e_i \rangle$,
corresponding to the square of the length (or norm) of
the respective projection vector of $\vert  \rho \rangle$ onto  $\vert  e_i \rangle$.
For complex vector spaces one has to take the absolute square of the scalar product;
that is, $f_\rho (  \vert  e_i \rangle   ) = \vert \langle  e_i \vert  \rho \rangle \vert ^2$.

Pointedly stated, from this point of view the probabilities $f_\rho (  \vert  e_i \rangle   )$
are just the (absolute) squares of the coordinates
of a unit vector  $\vert \rho \rangle$ with respect to some orthonormal basis $\{  \vert  e_i \rangle   \}$,
representable by the square $\vert \langle  e_i \vert  \rho \rangle \vert ^2$ of the length of the vector projections of
  $\vert \rho \rangle$ onto the basis vectors   $\vert e_i \rangle$.
The squares come in because the absolute values of the individual components do not add up to one; but their squares do.
These considerations apply to Hilbert spaces of any, including two, finite dimensions.
In this non-general, {\it ad hoc} sense the Born rule for a system in a pure state and an elementary proposition observable
(quantum encodable by a one-dimensional projection operator) can be motivated by the requirement of subclassicality for arbitrary finite dimensional Hilbert space.

Note that it is possible to generate ``Boole-Bell type inequalities (sort of)''
if one is willing to {\em abandon subclassicality}.
That is, suppose one is willing to accept that, within any particular context
mutually excluding observables are not mutually exclusive any longer.
In particular, one could consider two-valued measures in which all or some or none
of the atoms acquire the value zero or one (with subclassicality, the two-valued measure
is one at only a single atom; all other atoms have measure zero).
With these assumptions one can, for every context,
define a ``correlation observable'' as the product of the (non-subclassical)
measures of all the atoms in this context.
For instance, for any particular $i$'th context $C_i$ with atoms
$a_{i,1}, \ldots , a_{i,n}$; then the ``joint probabilities'' $P_i$ or ``joint expectations'' $E_i$ of a single context $C_i$ take on the values
\begin{equation}
\begin{split}
P_i= \prod_{j=1}^n v(a_{i,j})=
v(a_{i,1})\cdots v(a_{i,n}),\\
E_i= \prod_{j=1}^n \left[1-2v(a_{i,j})\right]=
\left[1-2v(a_{i,1})\right]\cdots \left[1-2v(a_{i,n})\right].
\end{split}
\end{equation}
A geometric interpretation in terms of
convex correlation polytopes is then straightforward -- the tuples representing the edges of the polytopes
are obtained by the enumeration of the ``joint probabilities'' $P_i$ or the ``joint expectations'' $E_i$
for all the involved contexts $C_i$.

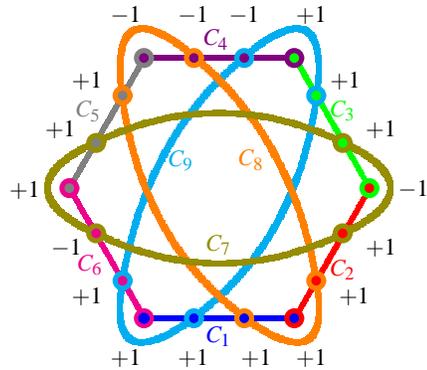
\begin{figure}
\begin{center}
\begin{tabular}{c}
\unitlength .4mm 
\allinethickness{2pt} 
\ifx\plotpoint\undefined\newsavebox{\plotpoint}\fi 
\begin{picture}(134.09,130)(0,-2)
\multiput(86.39,101.96)(.119617225,-.208133971){209}{{\color{green}\line(0,-1){0.208133971}}}
\multiput(86.39,14.96)(.119617225,.208133971){209}{{\color{red}\line(0,1){0.208133971}}}
\multiput(36.47,101.96)(-.119617225,-.208133971){209}{{\color{gray}\line(0,-1){0.208133971}}}
\multiput(36.47,14.96)(-.119617225,.208133971){209}{{\color{magenta}\line(0,1){0.208133971}}}
\color{blue}\put(86.39,15.21){\color{blue}\line(-1,0){50}}
\put(86.39,101.71){\color{violet}\line(-1,0){50}}
\color{cyan}
\qbezier(29.2,27.73)(23.55,-5.86)(52.99,15.24)
\qbezier(29.2,27.88)(36.93,75)(69.63,101.91)
\qbezier(52.69,15.24)(87.47,40.96)(93.72,89.27)
\qbezier(93.72,89.27)(98.4,125.99)(69.49,102.06)
\color{orange}
\qbezier(93.57,27.73)(99.22,-5.86)(69.78,15.24)
\qbezier(93.57,27.88)(85.84,75)(53.13,101.91)
\qbezier(70.08,15.24)(35.3,40.96)(29.05,89.27)
\qbezier(29.05,89.27)(24.37,125.99)(53.28,102.06)
\color{olive}
\qbezier(20.15,73.72)(-11.67,58.52)(20.15,43.31)
\qbezier(20.33,73.72)(61.34,93.16)(102.36,73.72)
\qbezier(102.36,73.72)(134.09,58.52)(102.53,43.31)
\qbezier(102.53,43.31)(60.99,23.43)(20.15,43.49)
\put(36.34,15.16){\color{magenta}\circle{5}}
\put(36.34,15.16){\color{blue}\circle{2}}
\put(52.99,15.16){\color{blue}\circle{2}}
\put(52.99,15.16){\color{cyan}\circle{5}}
\put(69.68,15.16){\color{blue}\circle{2}}
\put(69.68,15.16){\color{orange}\circle{5}}
\put(86.28,15.16){\color{blue}\circle{2}}
\put(86.28,15.16){\color{red}\circle{5}}
\put(93.53,27.71){\color{red}\circle{2}}
\put(93.53,27.71){\color{orange}\circle{5}}
\put(102.37,43.44){\color{red}\circle{2}}
\put(102.37,43.44){\color{olive}\circle{5}}
\put(111.21,58.45){\color{red}\circle{2}}
\color{green}\put(111.21,58.45){\circle{5}}
\put(102.37,73.47){\color{green}\circle{2}}
\put(102.37,73.47){\color{olive}\circle{5}}
\put(93.53,89.21){\color{green}\circle{2}}
\put(93.53,89.21){\color{cyan}\circle{5}}
\put(86.28,101.76){\color{green}\circle{2}}
\put(86.28,101.76){\color{violet}\circle{5}}
\put(69.68,101.76){\color{violet}\circle{2}}
\put(69.68,101.76){\color{cyan}\circle{5}}
\put(52.99,101.76){\color{violet}\circle{2}}
\put(52.99,101.76){\color{orange}\circle{5}}
\put(36.34,101.76){\color{violet}\circle{2}}
\put(36.34,101.76){\color{gray}\circle{5}}
\put(29.24,89.21){\color{gray}\circle{2}}
\put(29.24,89.21){\color{orange}\circle{5}}
\put(20.4,73.47){\color{gray}\circle{2}}
\put(20.4,73.47){\color{olive}\circle{5}}
\put(11.56,58.45){\color{gray}\circle{2}}
\put(11.56,58.45){\color{magenta}\circle{5}}

\put(20.4,43.44){\color{magenta}\circle{2}}
\put(20.4,43.44){\color{olive}\circle{5}}
\put(29.24,27.71){\color{magenta}\circle{2}}
\put(29.24,27.71){\color{cyan}\circle{5}}
{\color{black}
\put(30.41,116)   {\makebox(0,0)[cc]{$-1$}}
\put(30.41,2)     {\makebox(0,0)[cc]  {$+1$}}
\put(52.68,116)   {\makebox(0,0)[cc]{$-1$}}
\put(52.68,2)     {\makebox(0,0)[cc]   {$+1$}}
\put(91.93,116)   {\makebox(0,0)[cc] {$+1$}}
\put(91.93,2)     {\makebox(0,0)[cc]  {$+1$}}
\put(69.65,116)   {\makebox(0,0)[cc]{$-1$}}
\put(73.65,2)     {\makebox(0,0)[cc]   {$+1$}}
\put(103.24,94.22){\makebox(0,0)[cc]{$+1$}}
\put(17.45,94.22) {\makebox(0,0)[cc] {$+1$}}
\put(106.24,22.45){\makebox(0,0)[cc]{$+1$}}
\put(17.45,22.45) {\makebox(0,0)[cc] {$+1$}}
\put(115.13,77.96){\makebox(0,0)[cc]{$+1$}}
\put(8.55,77.96)  {\makebox(0,0)[cc]  {$+1$}}
\put(115.13,38.72){\makebox(0,0)[cc]{$+1$}}
\put(10.55,38.72) {\makebox(0,0)[cc] {$-1$}}
\put(120.92,57.98){\makebox(0,0)[l] {$-1$}}
\put(1.77,57.98)  {\makebox(0,0)[rc]  {$+1$}}
}
\put(61.341,9.192){\color{blue}\makebox(0,0)[cc]    {$C_1$}}
\put(102.53,31.355){\color{red}\makebox(0,0)[cc]   {$C_2$}}
\put(102.53,84.322){\color{green}\makebox(0,0)[cc]  {$C_3$}}
\put(60.457,108.01){\color{violet}\makebox(0,0)[cc] {$C_4$}}
\put(18.031,84.145){\color{gray}\makebox(0,0)[cc] {$C_5$}}
\put(18.561,33.057){\color{magenta}\makebox(0,0)[cc]{$C_6$}}
\put(61.341,39.774){\color{olive}\makebox(0,0)[cc]  {$C_7$}}
\put(72.124,67.882){\color{orange}\makebox(0,0)[cc] {$C_8$}}
\put(48.79,67.705){\color{cyan}\makebox(0,0)[cc]    {$C_9$}}
\end{picture}
\end{tabular}
\end{center}
\caption{(Color online) Orthogonality diagram of a finite subset
$C_1,\ldots ,C_9$ of the continuum of blocks or contexts embeddable in
four-dimensional real Hilbert space without a two-valued probability
measure~\cite{cabello-96};
with one of the 1152 non-admissible value assignments yielding the bound seven,
as derived in Ref~\cite{cabello:210401}.
In contrast, subclassicality would require that, within each one of the nine contexts, exactly one observable
would have value ``$-1$,'' and the other three observables would have the value ``$+1$.''
\label{2007-miracles-ksc}}
\end{figure}

For example, solving the hull problem for the ``correlation polytope''
of a system of observables introduced in Ref.~\cite{cabello-96}
and depicted in Fig.~\ref{2007-miracles-ksc}
yields, among 274 facet inequalities,
\begin{equation}
\begin{split}
0\le P_1\leq 1 \\
P_1 + 3\geq P_2 + P_6 +  P_7 + P_8 \\
P_1 + P_3 +  P_5 + 4\geq P_2 +  P_4 + P_6 +  P_7 + P_8 + P_9 \\
\ldots \\
-1 \leq  E_1 \leq 1, \\
E_1+7\geq E_2+E_3+E_4+E_5+E_6+E_7+E_8+E_9, \\
E_1+E_8+E_9+7\geq E_2+E_3+E_4+E_5+E_6+E_7, \\
E_1+E_6+E_7+E_8+E_9+7\geq E_2+E_3+E_4+E_5, \\
E_1+E_4+E_5+E_6+E_7+E_8+E_9+7\geq E_2+E_3, \\
E_1+E_2+E_3+E_4+E_5+E_6+E_7+E_8+E_9+7\geq 0
.
\end{split}
\label{2015-s-e11}
\end{equation}
The last bound has been introduced in Ref.~\cite{cabello:210401}.
It is violated both by
classical models (satisfying subclassicality)
as well as by quantum mechanics, because both cases obey subclassicality,
thereby rendering
the value ``$-1$'' for any  ``correlation observable''
$E_1, \ldots , E_9$
of all nine tightly intertwined contexts $C_1,\ldots , C_9$:
in each context, there is an odd number of ``$-1$''-factors.
For the sake of demonstration, Fig.~\ref{2007-miracles-ksc} also
explicitly enumerates one (of 1152 non-admissible, non-subclassical)
value assignments
yielding the bound seven.

However, note that the associated
observables, and also the two-valued measures and frame functions,
have been allowed to disrespect subclassicality;
because otherwise no two-valued measure exists.

Note also that similar calculations~\cite{pitowsky-89a,Pit-91,Pit-94,2000-poly}
for two- and three-partite correlations
do not suffer from a lack of subclassicality, since in an Einstein-Podolsky-Rosen setup,
the observables entering as factors in the product -- coming from different particles -- are independent (therefore justifying
multiplication of single-particle probabilities and expectations), and not part of a one and the same single-particle context.

\section{Discussion}

We have discussed ``bizarre'' structures of observables
and have considered classical,
quantum
and other, more ``bizarre''
probability measures on them.
Thereby we have mostly assumed
subclassicality, which stands for additivity within contexts, formalized by frame functions
as well as admissibility~\cite{Gleason,r:dvur-93,pitowsky:218,Peres-expTest-Glea}.

From all of this one might conclude a simple lesson:
in non-Boolean empirical structures which
allow both a quantum as well as a quasi-classical representation
(rendering a homeomorphic embedding into some larger Boolean algebra)
the predictions from quantum and classical probabilities
(rendered by the convex combination of two-valued measures)
may be different.
Which ones are realized depends on the nature of the system
(e.g. quasi-classical generalized urn models or finite automata,
or quantum states of orthohelium~\cite{kochen1}) involved.

Such structures may also allow (non-dispersive) probabilities and predictions
which can neither be realized by (quasi-) classical nor be quantized systems.
Stated pointedly:
even if one assumes subclassicality -- that is, the validity of classical predictions
within contexts in the form of maximal subsets of observables which are mutually co-measurable --
in general (i.e. in non-Boolean cases) the structure of observables does not
determinate the probabilities completely.

Finally, let us speculate that if we were living in a computable universe
capable of universal computation,
then universality would imply
we could see the types of collections of observables sketched in Figs.~\ref{2015-s-f1}-\ref{2015-s-f2};
at least if some (superselection) rule would not prohibit the occurrence
of such propositional structures.
Why do we not observe them?
Maybe we have not looked closely enough,
or maybe the Universe is not entirely
``universal'' in terms of fundamental  phenomenology.

I personally have a rather simple stance towards these issues,
which comes out of my inclinations~\cite{svozil-2013-omelette}
towards
{\em ``The Church of the larger Hilbert space.''}
I believe that Dirac~\cite{dirac} and von Neumann~\cite{v-neumann-49}
had it all right -- alas in a surprising, literal way.
The quantum universe appears to be the geometry of
linear vector space equipped with a scalar product (projections).
From this point of view, all those bizarre structures of observables
and prediction probabilities do not show up just because,
after all, our operationally accessible universe,
at least on the most fundamental level,
has to be understood in purely geometric terms,
thereby disallowing some algebraic possibilities.
This may be similar to the non-maximal violation of certain Boole-Bell type conditions of possible experience.

\begin{acknowledgments}
This research has been partly supported by FP7-PEOPLE-2010-IRSES-269151-RANPHYS.
I gratefully acknowledge advice from Komei Fukuda with the {\tt cddlib} package,
as well as Alexander Svozil for his help installing it.
I am also indebted to Alastair A. Abbot for a critical reading of the manuscript,
as well as for suggesting improvements.
\end{acknowledgments}

\ifws

\else


%

\fi
\end{document}